\documentclass[fleqn,usenatbib]{mnras}

\usepackage{newtxtext,newtxmath}

\usepackage[T1]{fontenc}
\usepackage{ae,aecompl}

\usepackage{caption}
\usepackage{subcaption}

\usepackage{graphicx}	
\usepackage{amsmath}	

\usepackage[figure,figure*]{hypcap}
\usepackage{float}

\newcommand{\comm}[1]{}

\defcitealias{2019MNRAS.490.1539R}{R19}

\title[B/PS bulges in DESI Legacy edge-on galaxies]{B/PS bulges in DESI Legacy edge-on galaxies I: Sample building}

\author[A. A. Marchuk et al.]{
Alexander A. Marchuk,$^{1}$\thanks{E-mail:a.marchuk@spbu.ru}
Anton A. Smirnov,$^{1}$
Natalia Y. Sotnikova,$^{2}$
Dmitriy A. Bunakalya,$^{2}$
\newauthor
Sergey S. Savchenko,$^{1,2,3}$
Vladimir P. Reshetnikov,$^{2}$
Pavel A. Usachev,$^{1,2,3}$
Iliya S. Tikhonenko,$^{2}$
\newauthor
Viktor D. Zozulia,$^{2}$
Daria A. Zakharova$^{2}$
\\
$^{1}$Central Astronomical Observatory at Pulkovo of RAS, Pulkovskoye Chaussee 65/1, 196140 St. Petersburg, Russia\\
$^{2}$St.Petersburg State University, 7/9 Universitetskaya nab., St.Petersburg, 199034, Russia\\
$^{3}$Special Astrophysical Observatory, Russian Academy of Sciences, 369167 Nizhnij Arkhyz, Russia\\
}

\date{Accepted XXX. Received YYY; in original form ZZZ}

\pubyear{2020}

\begin{document}
\label{firstpage}
\pagerange{\pageref{firstpage}--\pageref{lastpage}}
\maketitle

\begin{abstract}
We present the biggest up-to-date sample of edge-on galaxies with B/PS bulges and X-structures. 
The sample was prepared using images from the DESI Legacy catalogue and contains about 2000 galaxies. 
To find suitable candidates in catalogue, we made the assumption that the residues (original images minus model) of galaxies with B/PS bulges should exhibit a characteristic X-shape. Galaxies with such features were selected by eye and then used as input data for a neural network training, which was applied to a bigger sample of edge-on galaxies. 
Using the available data and the photometric models from the literature, we investigated the observational and statistical properties of the sample created. 
Comparing the $B/D$ ratios for galaxies with and without B/PS bulges, we found that the $B/D$ ratio for galaxies from our sample is statistically higher, with typical values in the range $\approx 0.2-0.5$ depending on the decomposition procedure. 
We studied how the opening angles $\varphi$ of the X-structure and the length of its rays are distributed in the formed sample and found them to be consistent with previous measurements and predictions from $N$-body models, e.g. $\varphi \gtrsim 25~\deg$, but measured here for a much larger number of galaxies. 
We found a sharp increase in the B/PS bulge fraction for stellar masses~$\log M_{\star} \gtrsim 10.4$, but for edge-on galaxies, which complements the results of previous works.
The sample can be used in future work to test various bar models and their relationship with B/PS bulges, as well as to study their stability and evolution. \end{abstract}




\label{lastpage}

\section{Introduction }


\par
Boxy/peanut-shaped (B/PS) bulges are thought to be thick parts of bars observed in edge-on or close to edge-on galaxies~\citep{Bertola_Capaccioli1977,Combes_Sanders1981,Kormendy_Illingworth1982,Combes_etal1990,Pfenniger_Friedli1991,Raha_etal1991,Kuijken_Merrifield1995, Bureau_Freeman1999, Merrifield_Kuijken1999, Veilleux_etal1999, Athanassoula_Misiriotis2002,Oneil_Dubinski2003, Chung_Bureau2004, Athanassoula2005,MartinezValpuesta_etal2006,Debattista_etal2006}. Since they are parts of bars seen in projection, they are also quite common inhabitants of galaxies in local Universe and can be found in about 20\%-40\% of disc galaxies~\citep{Shaw1987,deSouza_DosAnjos1987,Lutticke_etal2000,Erwin_Debattista2013,Yoshino_Yamauchi2015,Erwin_Debattista2017}. This fraction increases significantly, up to about 70\%, if one takes into account various projection effects~\citep{Kruk_etal2019}.
\par
B/PS bulges are formed during the secular evolution of the galaxy and this stage plays an important role in the formation of the structure of galaxies as a whole. Moreover, the structure of the B/PS bulge depicts the history of the underlying galaxy as the properties of B/PS bulges are determined by the properties of their host galaxies.
This was already proved in many numerical studies~\citep{Athanassoula_Misiriotis2002, Oneil_Dubinski2003, Athanassoula2005, MartinezValpuesta_etal2006, Debattista_etal2006, Wozniak_Michel-Dansac2009, Saha_Gerhard2013, Fragkoudi_etal2017, Smirnov_Sotnikova2018,Parul_etal2020,Sellwood_Gerhard2020}. For example, it was found that if an additional central component (like a classical bulge) presents in a galaxy then a B/PS bulge tends to grow symmetrically without a clear buckling phase~\citep{Smirnov_Sotnikova2018, Smirnov_Sotnikova2019, Sellwood_Gerhard2020}. B/PS bulges usually have one bright photometric feature, the so-called X-structures. The parameters of the X-structures are also related to the parameters of the galaxy itself. For example, depending on the initial conditions in the stellar disc, 
B/PS bulge can have double X-structures~\citep{Debattista_etal2017, Smirnov_Sotnikova2018, Parul_etal2020, Ciambur_etal2021}. \cite{Smirnov_Sotnikova2018} also found that galaxies with a significant contribution of the dark matter to the overall gravitational potential should have more flattened X-structures and, therefore, more flattened B/PS bulges.
\par
The aforementioned dependencies have been established for model systems, while they have not been clearly confirmed for real galaxies.
For real galaxies, the study of B/PS bulges is more complicated compared to models due to various projection effects (change of B/PS observable properties with the bar viewing angle or the disc inclination), an unavoidable mixing with other components (bulges of other types, discs, etc.), and, of course, dust attenuation. 
That is why the detailed investigations of the B/PS bulges were carried out only for small samples of galaxies~(\cite{Ciambur_Graham2016,Laurikainen_Salo2017,Savchenko_etal2017,Smirnov_Savchenko2020}, in total, there are about 40 edge-on galaxies in these lists). This substantially limits our understanding of how B/PS bulges live and evolve in real galaxies.
\par
It would be possible to enlarge the sample of galaxies with B/PS bulges by including galaxies observed almost face-on or at intermediate inclinations~\citep{Erwin_Debattista2013,Mendez-Abreu_etal2018, Kruk_etal2019}. In such galaxies, the thick part of the bar (the B/PS bulge) manifests itself in the disc plane as the boxy-like isophotes in the central area of the bar. But it seems that such a feature is not a mandatory one. At least,~\cite{Laurikainen_Salo2016} with reference to~\cite{Athanassoula_etal2015} clearly stated that the B/PS bulges can form in such a way that the isophotes profile in the disc plane remains ``fairly round''. Moreover, barlenses with round isophotes and B/PS bulges are perceived by some authors as the same entity but observed at different inclinations angles~\citep{Laurikainen_Salo2017}, although this concept has met some critique~\citep{Tikhonenko_etal2021}. Due to the lack of agreement on how to detect B/PS bulges in galaxies at intermediate inclinations, we will take a ``safer'' approach and discuss B/PS bulges and their properties mainly in the context of edge-on or close to edge-on galaxies.
\par 
The main goal that we set in this article is to significantly expand the existing list of edge-on galaxies with clearly detectable and bright B/PS bulges suitable for detailed photometric analysis. To find objects, we use the most modern methods - the Machine Learning methods and artificial neural networks.
\par
Rapidly developed usage of various Machine Learning methods in Astronomy (see \citealt{Baron2019}) and especially of convolutional neural networks (CNN) for images \citep{LeCun1990} makes classification of large collections of galaxies possible. Besides other commonly applied Machine Learning (ML) supervised algorithms, CNN does not demand a sophisticated process of feature creation and thus can solve a variety of tasks of different nature. In addition, often artificial neural networks (ANNs) and CNN in particular greatly outperform traditional algorithms. These models found application in such astronomy tasks as $B/T$ prediction \citep{2021MNRAS.506.3313G}, redshift estimation \citep{2019A&A...621A..26P}, star formation rate estimation \citep{2020MNRAS.493.4808S}, prediction of galaxies shapes \citep{2019MNRAS.489.4847R} and morphology \citep{2021arXiv211012735W}, galaxy inclination angle estimation \citep{2020MNRAS.497.3323P}, stellar spectral classification \citep{2020MNRAS.491.2280S}, detection of anomalies in astronomy data \citep{2021A&C....3600481L}, and numerous others. It is also worth noting works of particular interest here, where some components of galaxies such as bars~\citep{2018MNRAS.477..894A}, Sersic profiles~\citep{2021arXiv211105434L} and individual star clusters \citep{2020AJ....160..264B} were detected.
\par
We aim to identify new candidates in the Catalog of Edge-on disc Galaxies from SDSS (EGIS)~\citep{Bizyaev_etal2014} and in the Edge-on Galaxies in the Pan-STARRS survey (EGIPS)~(\citealt{2022MNRAS.tmp..256M}) using images from Dark Energy Spectroscopic Instrument (DESI) Legacy catalogue~\citep{Dey2018}. In earlier works~\citep{Shaw1987,deSouza_DosAnjos1987,Lutticke_etal2000,Erwin_Debattista2013,Yoshino_Yamauchi2015,Erwin_Debattista2017}, the existence of a B/PS bulge in a particular galaxy was verified simply by a visual inspection. With the aim to acquire as many candidates as possible, we decided to automatise the process of distinguishing B/PS bulges in this work. As we mention earlier, this can be done with a new but already well-established method of image processing --- artificial neural networks, one of the original purposes of which is to classify different types of objects. 
\par
This is the first work in the series and hence ``I'' in the title, 
where we describe our approach to distinguishing galaxies with B/PS bulges, as well as the main properties of the identified sample of galaxies. Detailed properties of B/PS bulges and results of more rigorous photometric analysis will be presented in the next paper.
\par
The structure of this paper is as follows.
\par 
In Section~\ref{sec:sample_building}, we describe the original data sets (EGIS, EGIPS and DESI Legacy catalogues) that we use to identify the galaxies with B/PS bulges in this study. The Section also contains various technical details of our approach, including preparing a training sample based on EGIS candidates, setting up our ANN, and manual processing candidates in the Zooniverse framework. 
\par
In Section~\ref{sec:gen_prop}, we examine the basic properties of our sample based on data available in the literature. This includes the distributions over the distances, stellar magnitudes, galaxy types, as well as various size and shape metrics. In subsection~\ref{sec:validation}, we present several tests of our sample to further validate our approach. In subsection~\ref{sec:interestingobj}, we present several examples of galaxies with large and easily detectable B/PS bulges that have been identified in the present study and, at the same time, have not previously been studied in relevant studies (except one).
\par
Section~\ref{sec:prelim_analysis} presents the results of the bulge-to-total contribution analysis based on photometric models available in the literature as well as the results of the analysis of the specific properties of the X-structures (opening angle, linear size) obtained on the basis of visual inspection of the residues.  Subsection~\ref{sec:stellarmasses} is devoted to the study of the dependence of the B/PS bulge frequencies on the stellar mass, which was studied in detail in some previous works.
\par 
In Section~\ref{sec:conclusion}, we summarise our results.

\section{Sample building} 
\label{sec:sample_building}

Manual construction of a large sample of galaxies with B/PS bulges is difficult for several reasons. First, collecting enough data would require visual  inspection of tens of thousands of images for each person, which is a huge amount of work. Secondly, even for edge-on galaxies, B/PS components are usually faint compared to the disc, and it is difficult to discern such low-contrast features by eye in some cases. This factor leads to human errors and results in each image having to be processed by multiple researchers, again increasing the overall workload. Simple automated image processing techniques based on Computer Vision methods usually require manual manipulations, and are not reliable enough, and, as far as we know, have not yet been applied to solve the problem under consideration.
\par
These are the main reasons why such a sample has not been built before. The largest up-to-date sample of B/PS bulges viewed edge-on was collected manually in \citet{Yoshino_Yamauchi2015} and contains less than 300 objects. Fortunately, recent advances in observation and automatic image classification address both of these problems.
In this section, we describe in detail how we collect a sample of B/PS bulges in edge-on galaxies using ANN.

\subsection{Data description}
\label{sec:data_description}  

To automate the search of B/PS bulges by training the machine to find suitable galaxies, we need two different sets of objects, namely a training set divided into two classes and a lot of unclassified data. The training set includes galaxies with visible B/PS bulges, which will be used as positive examples for the ANN, and galaxies with a high probability of missing the B/PS feature, which will be used as negative examples in the training classification. We also need a set of objects to apply the trained model, which will hopefully give us the desired large sample of galaxies with B/PS bulges to study their properties. This last set of galaxies should be large enough and homogeneous with the training data, that is, should contain objects very similar to those already shown by the ANN. 
\par
In this paper, we focus on edge-on galaxies, since the B/PS bulges are best seen when the inclination of the disc is close to $90\deg$. The training set can only be built manually by eye, and for this task it is better to have data already prepared. Fortunately, we have two large samples of edge-on galaxies that can be used for both training and classification (see below). Moreover, some initial steps in this direction have already been made in \citet{Savchenko_etal2017}, where the authors manually selected 151 galaxies with clear B/PS bulges. Finally, edge-on galaxies with and without B/PS bulges have been studied for a long time in our group  \citep{Bizyaev_etal2014,Savchenko_etal2017,Mosenkov2015,Smirnov_Savchenko2020,edgeon1,edgeon2,edgeon3}.  
\par

All images to be used in this work are from the DESI Legacy Imaging Surveys DR8 \citep{Dey2018}. It consists of three public projects from different telescopes and collectively covers $\approx 14000 \deg^2$ in both the Northern and Southern celestial hemispheres (see contours on Fig.~\ref{fig:skypos}, the sky coverage is approximately bounded by -18 deg < $\delta$ < +84 deg). The survey observations are carried out in the $g, r, z$ bands and by 1-2 magnitudes deeper than SDSS \citep{2003AJ....126.2081A}, which makes it possible to reveal weaker details that were not previously observed. The survey also contains the results of photometric decompositions for all objects using an automated probabilistic software called \texttt{TRACTOR} \citep{tractor}. Models in \texttt{TRACTOR} are axisymmetric and can include Sersic and exponential profiles, but do not include more complex structures such as B/PS bulges. Thus, after  subtracting the model from the original image, we obtain a more contrast and visible residue image of the X-structure. As already mentioned, the X-structure is the most striking feature of the B/PS bulge, and it is its presence in the residue image that indicates the presence of the B/PS bulge in a galaxy. Availability of residual images with X-structures in DESI Legacy allows us for the first time to massively explore the faint parts of these galaxies, such as B/PS bulges, and this is the reason why for all images of galaxies below we will use the residuals layer of the survey, which is the original galaxy after the subtraction of the model (see examples in Figures in Sec.~\ref{sec:voting}). It should be noted, however, that the resulting models are not always ideal and, among other problems, it may be difficult to assign components to a particular galaxy due to probabilistic approach of~\texttt{TRACTOR}.
\par
We used the Edge-on Galaxies in SDSS (EGIS) catalogue \citep{Bizyaev_etal2014} to create training data. The catalogue includes 5747 edge-on galaxies, selected from SDSS DR8 in the $g,r,i$ bands, and is complete for galaxies with a major axis greater than 28~arcsec. The authors additionally checked that the galaxies in the catalogue are distributed over all Hubble morphological types, and also performed 1D and 3D decompositions of their discs. From 2021 large enough galaxies from EGIS, \citet{Savchenko_etal2017} selected 151 objects with a clear presence of B/PS bulges, and 22 of them were used for the subsequent analysis, including careful decomposition and estimation of parameters of B/PS bulges. A similar manual examination by eye was performed in this study, but instead, all EGIS galaxies regardless of size were used and we checked the DESI Legacy residual images in the $r$ band instead of original SDSS images. Since in the training sample we want to create the most complete set of galaxies with B/PS bulges, when visually examining the residual image, we select not only clear and massive examples of X-structures, but also those dimmed by dust, faint, warped, etc. (see criteria of quality classes in Sec.~\ref{sec:voting} and examples therein). The presence of the X-shape in the selected galaxies have been independently confirmed by at least one of the first three authors. 
\par
Overall, this study results in the detection of significantly more B/PS bulges than previously found in SDSS data, supporting the idea that low contrast features in the original images are difficult to detect and that DESI Legacy is of better quality. 
The final training sample consists of 637 EGIS galaxies with B/PS bulges and 4502 without them. These six hundreds of positive examples already constitute a larger sampler than each of the previously known ones, but we aim to obtain an even larger sample in order to get more statistically reliable insights. It should be noted, that the training sample size is slightly smaller than the original EGIS size because not all the galaxies observed in SDSS lie in the part of the sky that intersects with that observed by DESI telescopes. For this reason, we find only 112 galaxies selected in \cite{Savchenko_etal2017} as positive examples. We do not analyse the properties of the training sample separately here and refer the reader to Sec.~\ref{sec:samplepprops}.
\par
As mentioned above, to apply the trained model, we need a second large uniform sample of edge-on galaxies. Fortunately, such a sample, called EGIPS (Edge-on Galaxies in the Pan-STARRS survey; \citealt{2022MNRAS.tmp..256M};\textcolor{blue}{Savchenko et al.,  in prep.}), has recently been created. In these works, the authors train an ANN and conduct a long automatic search for edge-on galaxies over the entire part of the sky, covered by Pan-STARRS DR2 observations in the $g,r,$ and $i$ bands, which were then manually checked by a group of individual researchers. As a result, a sample of 16551 edge-on or near edge-on galaxies was obtained, 13,048 of which are also available in the DESI Legacy survey.
The authors found their catalogue to be complete for the semi-major axis characteristic sizes larger than 6 arcsec (\citealt{2022MNRAS.tmp..256M}). All available in DESI footprint galaxies from EGIPS will be used as candidates for galaxies with B/PS bulges and used as input to the trained classifier. We note that a significant part of 3231 galaxies from the EGIS catalogue is included in EGIPS as a subsample. We should also mention that it was the success of EGIPS that inspired us to try the same approach for finding B/PS bulges.

\subsection{ANN} 

\begin{figure}
\includegraphics[width=0.95\columnwidth]{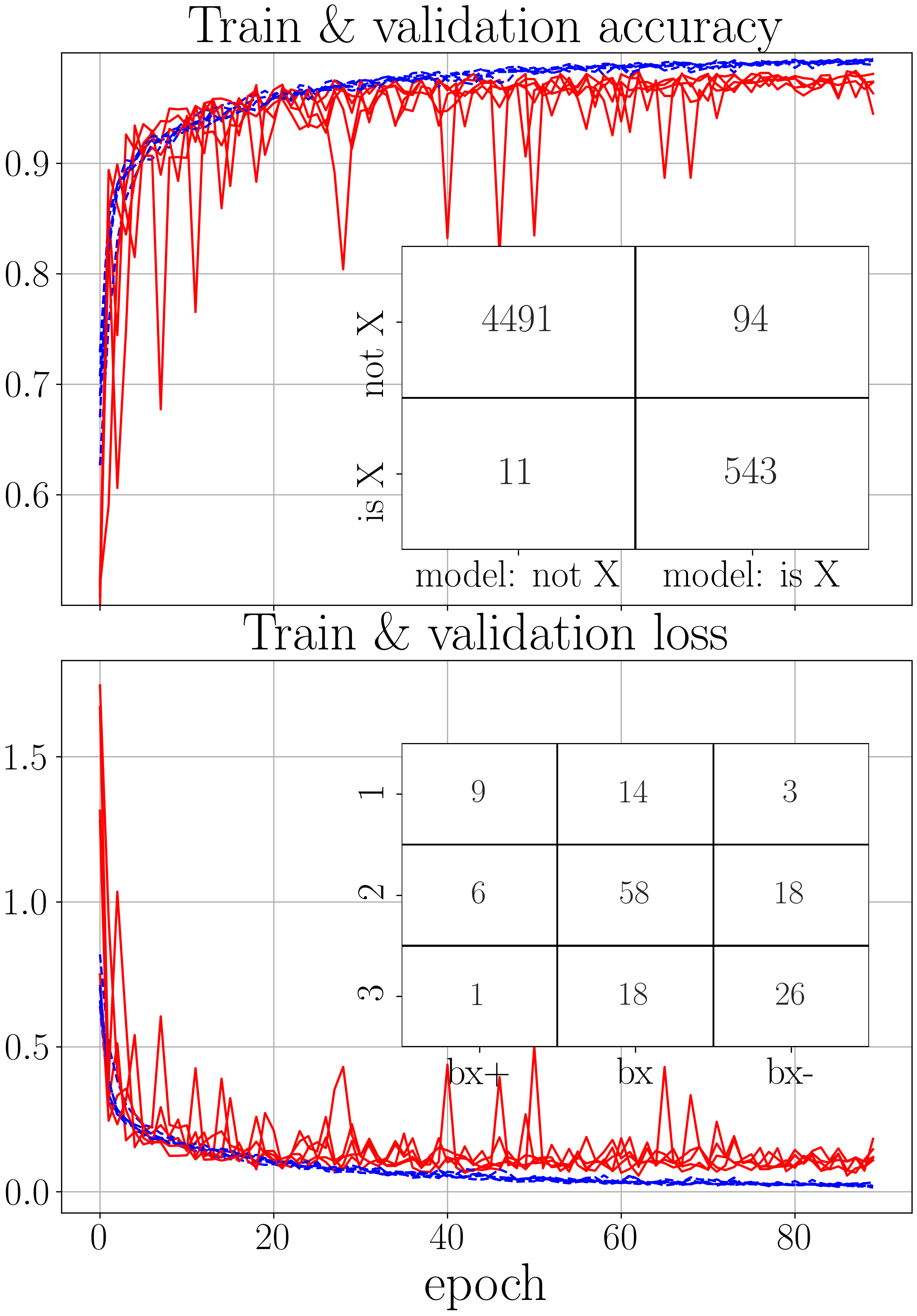}
\caption{The ANN train (blue) and validation (red) accuracy (top) and loss function values (bottom) as a function of epoch number for each of the five models. The inner image in the top plot represents the confusion matrix of the resulting model and the inner image in the bottom plot shows a comparison between our classes and the classes from \citet{Yoshino_Yamauchi2015}, see details in Sec.~\ref{sec:validation}.}
\label{fig:trainbyepoch}
\end{figure}

In everything related to ANN, we mainly follow \textcolor{blue}{Savchenko et al. (in prep.)} and refer the interested reader to find details in this work, including a better explanation of how CNN works and description of individual layers in ANN. Here we will highlight only the essential differences. The topology of ANN network used here is the same as presented in \textcolor{blue}{Savchenko et al. (in prep.)}, but adapted for a larger input data size, that is a larger galaxy image. In short, schematically, the network consists of three blocks, and each of them consists of two consequent convolution layers, followed by batch normalisation, max pooling and dropout layers\footnote{Description and links for each mentioned layer type can be found in \textcolor{blue}{Savchenko et al. (in prep.)}.}. The network is complemented by two fully connected layers and the final prediction whether the galaxy contains B/PS bulge is formed by the activation function \texttt{ReLU}, which is defined as the positive part of its argument, and the function \texttt{SoftMax} after it. 
\par
For each galaxy, we use the images in the $g,r,$ and $z$ bands as input for ANN. To create an ANN that focuses on important features of the image, that is, X-structure, and does not take into account that, for example, an edge-on galaxy can have different orientations, we perform some additional image processing before training. For each image, we centre it to the position of the galaxy centre. Then we use the positional angle (PA) of the disc major axis from the EGIS catalogue and rotate the image so that this axis becomes horizontal. It should be noted, that PA measurement is very reliable for edge-on galaxies. After that, we crop the image to a square of $1.5\times r_{\mathrm{eff}}$ size, where $r_{\mathrm{eff}}$ is the disc half-light radius, taken from \texttt{TRACTOR} data using \texttt{SHAPEEXP\_R} parameter for the exponential model, which is found as the closest one to the centre of the galaxy. Thus, after cropping, we focus only on the parts of interest located in the centre of the disc. The motivation behind such crop selection is rather empirical, but in practice it has been shown that the X-structure of the B/PS bulge rarely goes beyond the selected boundaries and is almost always fully represented in the image. We also limit the minimum image size to 15~arcsec. Finally, all images in all three bands, $g, r$, and $z$, were resized to a square of 64 × 64 pixels. Therefore, the input of ANN is a 3D array of the size $64\times64\times3$.
\par
The training sample, collected in Sec.~\ref{sec:data_description}, is small compared to that traditionally used as input to the ANN. It is also significantly imbalanced, where for every positive example we have almost 6-7 examples of negative class galaxies, i.e. those without visible X-structures. To address this issue, in each training experiment, described below, we randomly select positive or negative examples with equal probability until a certain total number of galaxies, in particular 10,000, is reached for each final model. Therefore, the ANN learns from a balanced dataset, where many negative examples are presented, and each galaxy with B/PS bulge appears multiple times. Despite the danger of overfitting, the used approach is sufficient to build a sample from the selected number of trained images, as the following analysis shows.
\par
The neural net was trained for 90 epochs, that is, the net was fed with the training dataset 90 times. We run the entire training multiple times, each time with a new randomly generated training set as described in previous paragraph, and the categorical cross-entropy were used as a loss function. It usually converges after 40-50 epochs without significant improvement thereafter, as Fig.~\ref{fig:trainbyepoch} suggests. Test data set for the loss estimation was generated randomly by selecting 20\% of the train part. These experiments show that the performance of individual models was unstable (note ``bumps'' in the red lines in Fig.~\ref{fig:trainbyepoch}) and that the models are probably slightly overfitted, which is natural due to the small number of positive examples. To fix the problem and get more stable results, we decide (again, following \textcolor{blue}{Savchenko et al. (in prep.)}, but for different reasons) to use an ensemble approach with voting. We trained 5 separate ANN models, each with its own slightly different training set, generated according to the rules described earlier. As a final decision on whether a B/PS bulge is present in the galaxy, we use the majority of the model votes, i.e. 3 out of 5 with positive decisions, and the probability threshold of the positive class in SoftMax was set to 0.95. However, experiments demonstrate that the stricter condition with 4 positive votes instead of 3 works better, resulting in fewer false positive examples and only in a small increase of false negative examples, so we decided to use this option. 
\par
Network performance for all five trained ensemble learners is shown in Fig.~\ref{fig:trainbyepoch}. It is clearly seen that the accuracy of all models reaches a plateau of about 0.96-0.98 and that they all show similar performance on validation sets with an accuracy of about 0.95-0.97. The number of correctly and incorrectly predicted galaxies presented in the form of the so-called confusion matrix,  which is shown in the inner subplot of Fig.~\ref{fig:trainbyepoch} for the complete EGIS data set and for the selected voting scheme. As can be seen, only eleven galaxies with the X-structure were not detected properly, contributing to the false negative class. These are galaxies with artefacts in the image or with a displaced centre. Thus, as follows from the confusion matrix, the ANN may miss some galaxies with the B/PS feature, but we can expect this fraction to be small. At the same time, about a hundred galaxies out of several thousand were detected as false positive. These galaxies do display some sign of the X-structure visible in the centre, often faint or distorted by dust. After careful consideration, we include 75 of them as examples of galaxies with B/PS bulges, which we missed at the first glance. This case again shows the complexity of manually processing large amounts of data and also the quality of the trained ANN. In general, we must conclude that even on such a small training sample, the performance of the resulting ANN ensemble is good and sufficient for the next step of searching for B/PS bulges in a larger dataset.
\par
All EGIPS images in the $g,r,$ and $z$ bands were prepared in the same way as the training set, where we get all needed parameters from the catalogue (\citealt{2022MNRAS.tmp..256M}). Then the ANN ensemble was applied to them with the same voting scheme as above. This led to the detection of 1828 candidates to galaxies with B/PS bulges from Pan-STARRS DR2 data. Before further analysis, a very simple test was performed to assess how good the ANN results were compared to manual ones. The first author visually checked all candidates and approved 1589 as correct, which is about 12\% of the original edge-on galaxy sample. We then select three times 100 random galaxies from the EGIPS sample and manually count the number of B/PS bulges in them. We found 14, 14 and, 18 objects, which is 15\% on average. This is larger than the fraction found by ANN, but not significantly. Obviously, the ANN cannot find some X-structures that humans can find, but the ANN's prediction for the B/PS fraction is very close to the estimates of the training data from EGIS, which is again 12\%. In any case, 4 positive votes out of 5 is a stricter condition than a simple majority of votes and  we can conclude that such a simple test shows ANN performance as reliable and good for the next investigation.
\par
To conclude the subsection on ANNs, we want to emphasize the following. 
Although ANN is a useful tool for finding suitable candidates, it is still not ideal because the trained set is limited and the detection rate is always less than 100\%. Hence, we still need the sample to be validated by the human eye in order to rule out false positives and be confident in forthcoming scientific insights. To eliminate human bias, correctly assess the performance of the model, identify problems and B/PS features, and form the final sample suitable for analysis, we conduct voting among all authors on the Zooniverse platform, which is described in the next subsection.

\begin{figure}
\includegraphics[width=0.95\columnwidth]{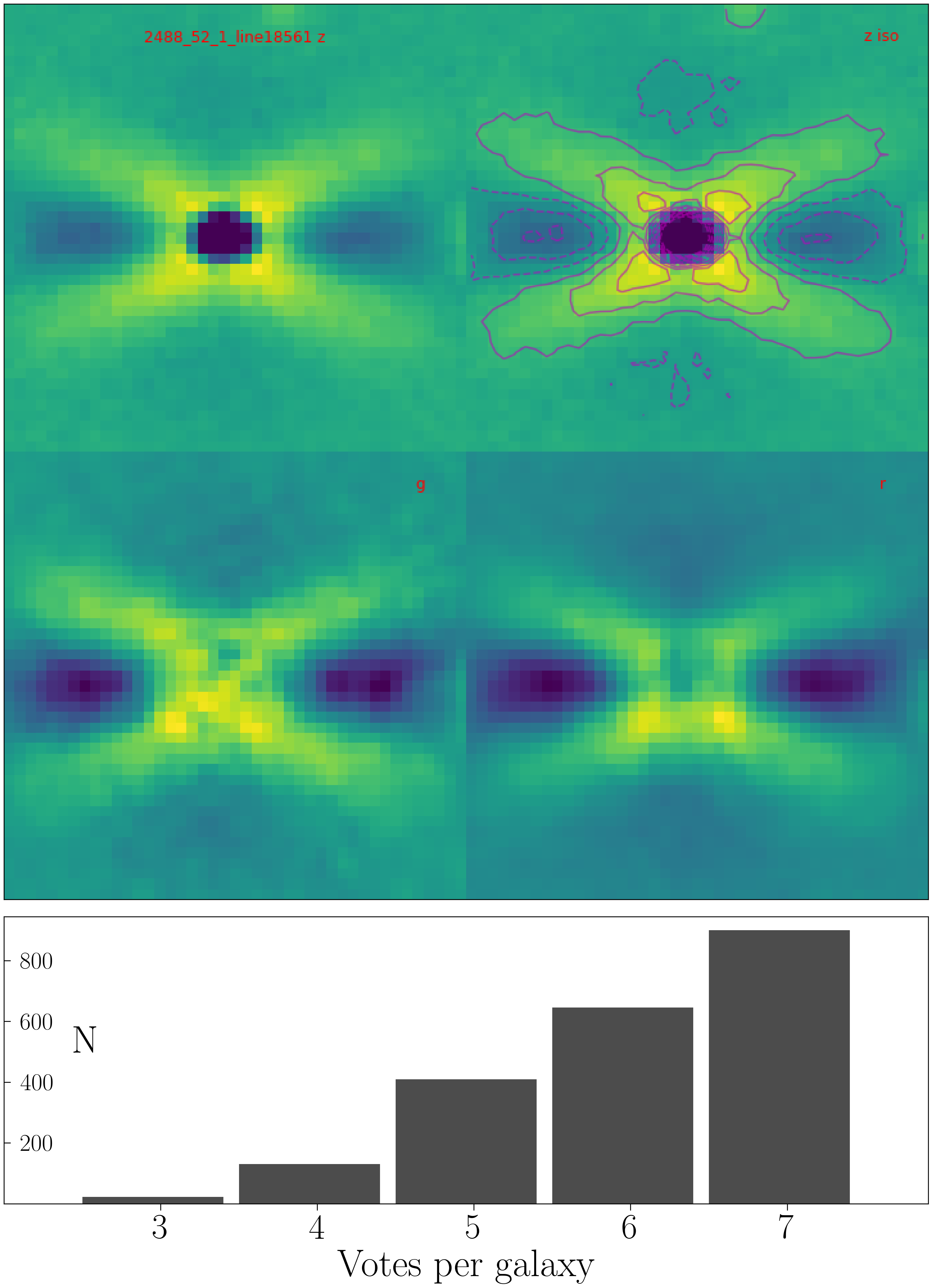}
\caption{The top plot demonstrates an example of a picture shown during Zooniverse voting and depicts the same galaxy in different bands, in the $z$ band with an overlaid galaxy identifier, in the $z$ band with isolines in the upper row, in the $g$ and $r$ bands in the bottom. The bottom subplot shows the distribution of collected votes. 
}
\label{fig:zooexample}
\end{figure}

\subsection{Voting}
\label{sec:voting} 

Unfortunately, despite the fact that the main goal was automatic B/PS detection, the obtained candidates in galaxies with B/PS bulges still need to be checked manually, because, firstly, the trained model may be wrong, and, secondly, not all candidates are suitable for further analysis. For this manual check, we create a set of five questions on Zoonverse\footnote{\url{https://www.zooniverse.org/}} web platform for citizen science. All authors participate in the responses, and for each candidate we collect at least three responses (see the histogram in Fig.~\ref{fig:zooexample}, and the median and mean are 6), which will be processed in detail in the next section and in the future papers.
\par
\begin{figure}
\includegraphics[width=0.95\columnwidth]{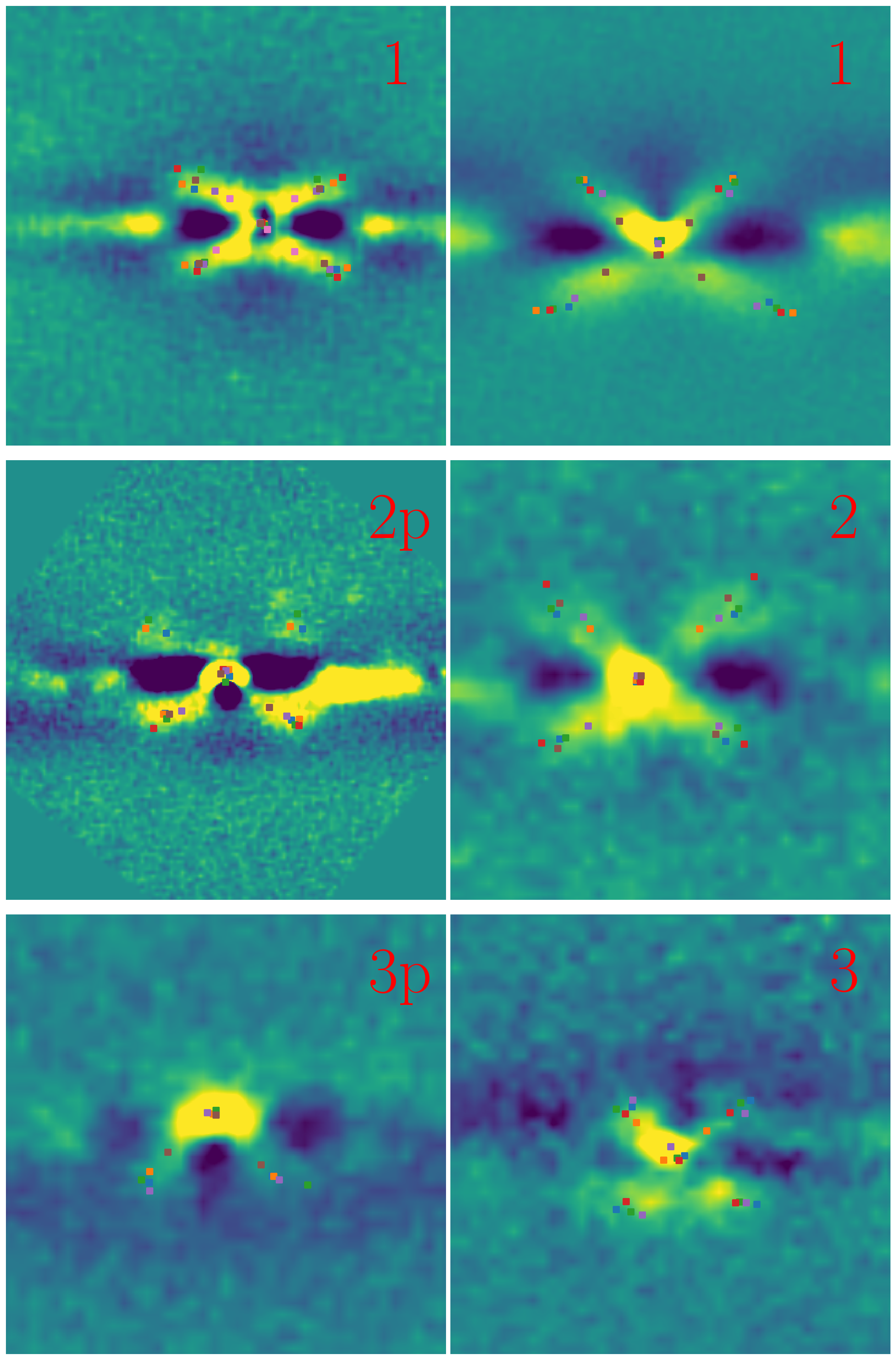}
\caption{Examples of galaxies of different quality classes, including galaxies with a partially visible X-structure (`p'). The images shown here are residues in the $z$ band that show clear presence of the X-structure. Note that some galaxies are not centred correctly. Dots of different colours show the positions of rays and centres marked by different participants.}
\label{fig:quality_examples}
\end{figure}

Let us briefly describe the proposed scheme of questions. For each candidate we show a grid with four pictures of a galaxy in $g,r,z$ bands plus in the $z$ band with isolines, see Fig.~\ref{fig:zooexample} as an example. All images are scaled using the IRAF zscale algorithm implemented in \texttt{astopy}~\citep{astropy1,astropy2}. The first question is whether it is a galaxy with a B/PS bulge or not, and if `yes', what is its quality. We consider quality as a subjective assessment, which is necessary because images with different noise and visibility of the X-structure are included in the same sample and may exhibit slightly different properties. We assess quality in three subjective classes, where class ``1'' consists of the best candidates and corresponds to clearly visible X-structures. Class ``2'' is when the overall quality or resolution is poorer and rays of the  X-structure are less bright. The last one, class ``3'', is when there is the X-structure in the image, but it is faint, very noisy, and difficult to distinguish. We naturally expect the galaxies of class ``1'' to be closer and brighter than the rest. We also introduce a subclass `p' (from `partial') for cases when only half of the four rays is visible due to the combined effect of the dust and a small disc inclination. Examples of galaxies for all classes can be found in Fig.~\ref{fig:quality_examples}. 
\par
The second question in the survey is about image quality. Participants may mark the cases where X-structures are not in the centre of the image or the major axis is not parallel to $x$-axis, or when the image contains artefact or a B/PS feature interferes with another object. In addition, each author can note if an X-structure is not visible in all photometric bands, whether it has poor resolution or is very weak. The third question concerns the dust line in the galaxy, whether it is clumpy and fuzzy, or too wide, or warped. The fourth question concerns the features of the X-structure, such as the asymmetry of its rays.
The last fifth question allowed participants manually put dots at the ends of the X-structure rays, which allowed us to roughly estimate the opening angles and X-structure size for these candidates (see Sec.~\ref{sec:angles} and \ref{sec:xsizes}).
\par
Questions were asked for images from candidates selected by ANN from Pan-STARRS DR2 data and also for objects in training set, manually selected from EGIS. After excluding intersected galaxies, the total size of the combined sample is 2123 objects. In this work, we do not analyse in detail the answers of the participants to all questions, except for the answers to the first (the X-structure quality) and the last (an X-structure opening angle) questions, leaving all the rest for a more sophisticated analysis in the future.

\section{General properties}
\label{sec:gen_prop}

\subsection{Voting evaluation}

In total, we collect 12935 votes for the sample under consideration. Each galaxy has at least 3 independent votes, and the histogram of votes number per galaxy is shown in Fig.~\ref{fig:zooexample}. 
\par
The final estimation in terms of X-structure quality was obtained in the following way. First, the quality class and the `partial' property `p' were
evaluated separately. We calculate `partial' probability as a fraction of votes for this option and assign index `p' to a particular galaxy if the mentioned probability is greater than 0.5. Inspired by Galaxy Zoo project \citep{2011MNRAS.410..166L}, 
we divide all votes into two categories of expert votes (first three authors of this publication) and participant votes. The motivation behind this decision is that the experts participated in the training sample building, and, thus, got more experience in the visual analysis of X-structures.
If all experts agree on some quality class, then we have high confidence in the decision and assign the galaxy to the corresponding class. Otherwise, we check whether the majority of experts (2 of 3 in our case) agree on some quality class and then compare their estimate with all other votes (including participants). We assign the galaxy to the corresponding class if the estimate of the majority of experts coincides with the estimate of the majority of the participants. After these two steps, more than two-thirds of galaxies in our sample are assigned to some quality class, and, in particular, 
almost all galaxies from class ``1''  are identified at this point. For all remaining cases, we decided to assign an average vote rounded to the closest class (the number 1, 2, or 3 was assigned for each class, respectively, and the "not X" option casts to 4).
\par
In the final sample, we ended up with 291 galaxies of class ``1'', 907 galaxies of class ``2'', and 727 galaxies of class ``3''. This is the biggest up-to-date collection of galaxies with B/PS bulges. 
It is also interesting to note that we have fewer galaxies of the third class than of the second, which is unusual compared to similar works of \citet{Lutticke_etal2000,Yoshino_Yamauchi2015}. This fact may be due to a natural human bias to select a median option or the fact that it is difficult for ANN to detect obscure cases. A more detailed comparison of our quality classes and~\cite{Yoshino_Yamauchi2015} is carried out below in Section~\ref{sec:validation}. We exclude 198 images for which we decide that they contain no B/PS bulges at all. Most of the excluded images contain X-shaped features in the centre, so it is not difficult to understand why ANN predict them as proper candidates. Most often, such a feature in the centre is due to a combination of spiral arms and the inclination of the disc plane. In several cases, it is a consequence of dust complex structure or because the galaxy is a distant one and has dim X-structure rays. We have 274 galaxies where a `partial' probability is bigger than the selected threshold of 0.5, among which are only 3 examples of the first class and 158 of the third. While the adopted scheme to evaluate classes may look slightly artificial, below we show that the properties of galaxies vary over classes in a physically consistent manner (e.g. Sec.~\ref{sec:BD} and Sec.~\ref{sec:angles}), thereby supporting the adopted scheme.



\begin{figure}
\includegraphics[width=0.95\columnwidth]{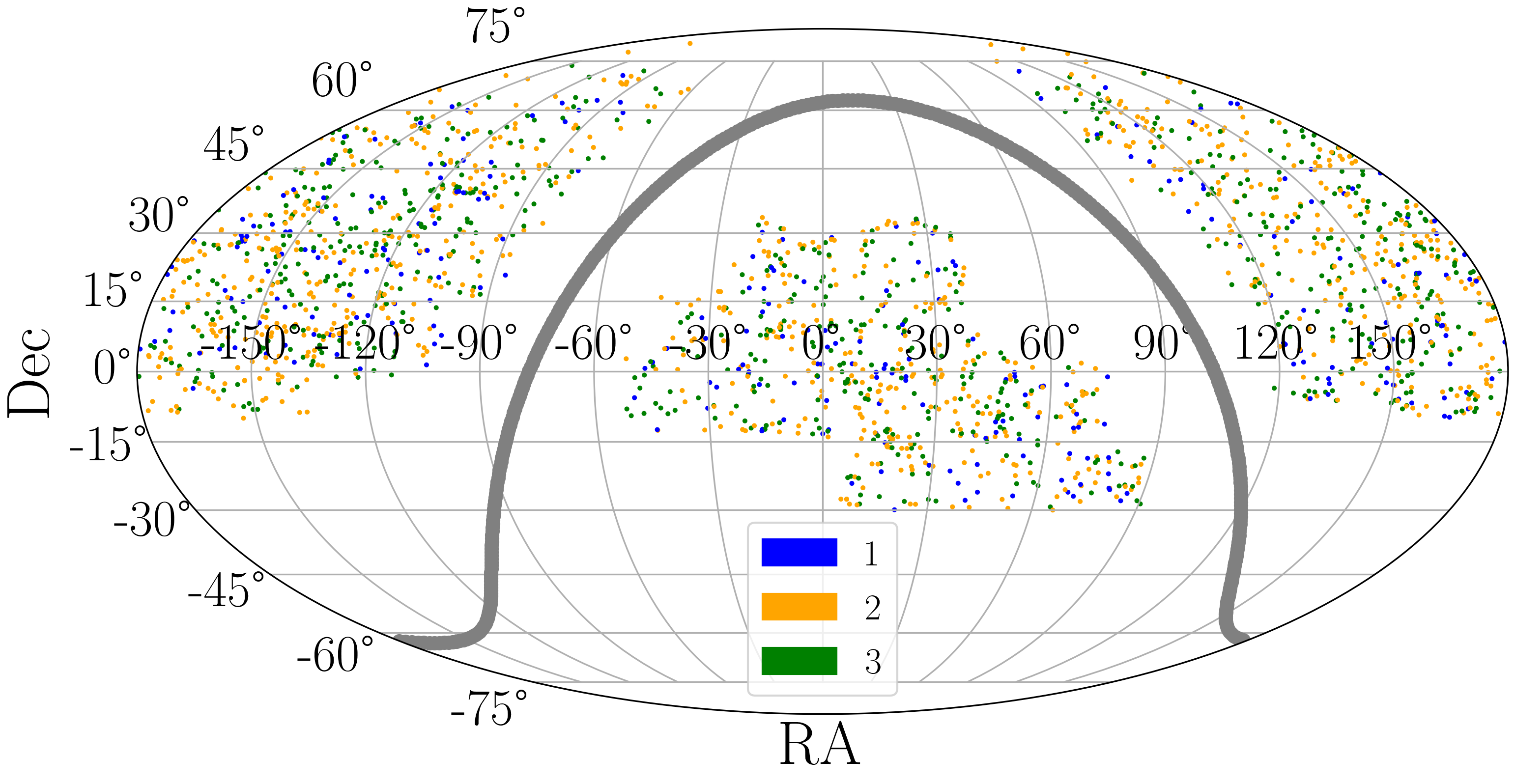}
\caption{Distribution of selected galaxies with B/PS bulges in the sky in equatorial coordinates. The colours represent galaxies of different classes, the grey line shows the plane of our Galaxy.}
\label{fig:skypos}
\end{figure}

\begin{figure}
\includegraphics[width=0.95\columnwidth]{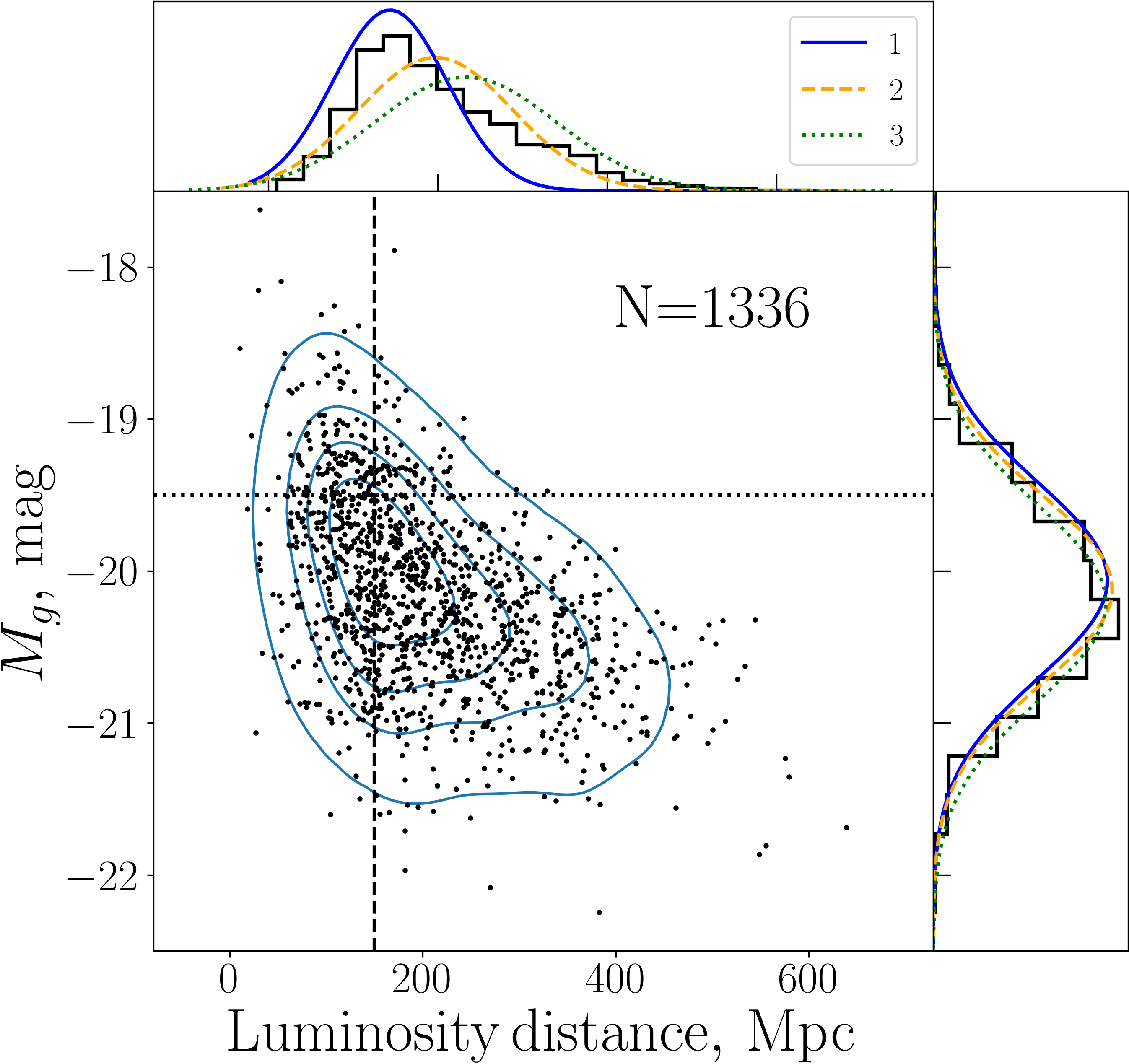}
\caption{The absolute stellar magnitude $M$ in the $g$ band versus the luminosity distance $D$. Panels on the sides show the distributions of both parameters for the entire sample (black histograms), the distributions for individual classes are shown by lines, normalized to the same area. Vertical and horizontal lines separate the two studied subsamples $S_{d<150}$ and $S_{M<-19.5}$ (see the main text for details).}
\label{fig:Mabs}
\end{figure}

\begin{figure*}
\includegraphics[width=1.95\columnwidth]{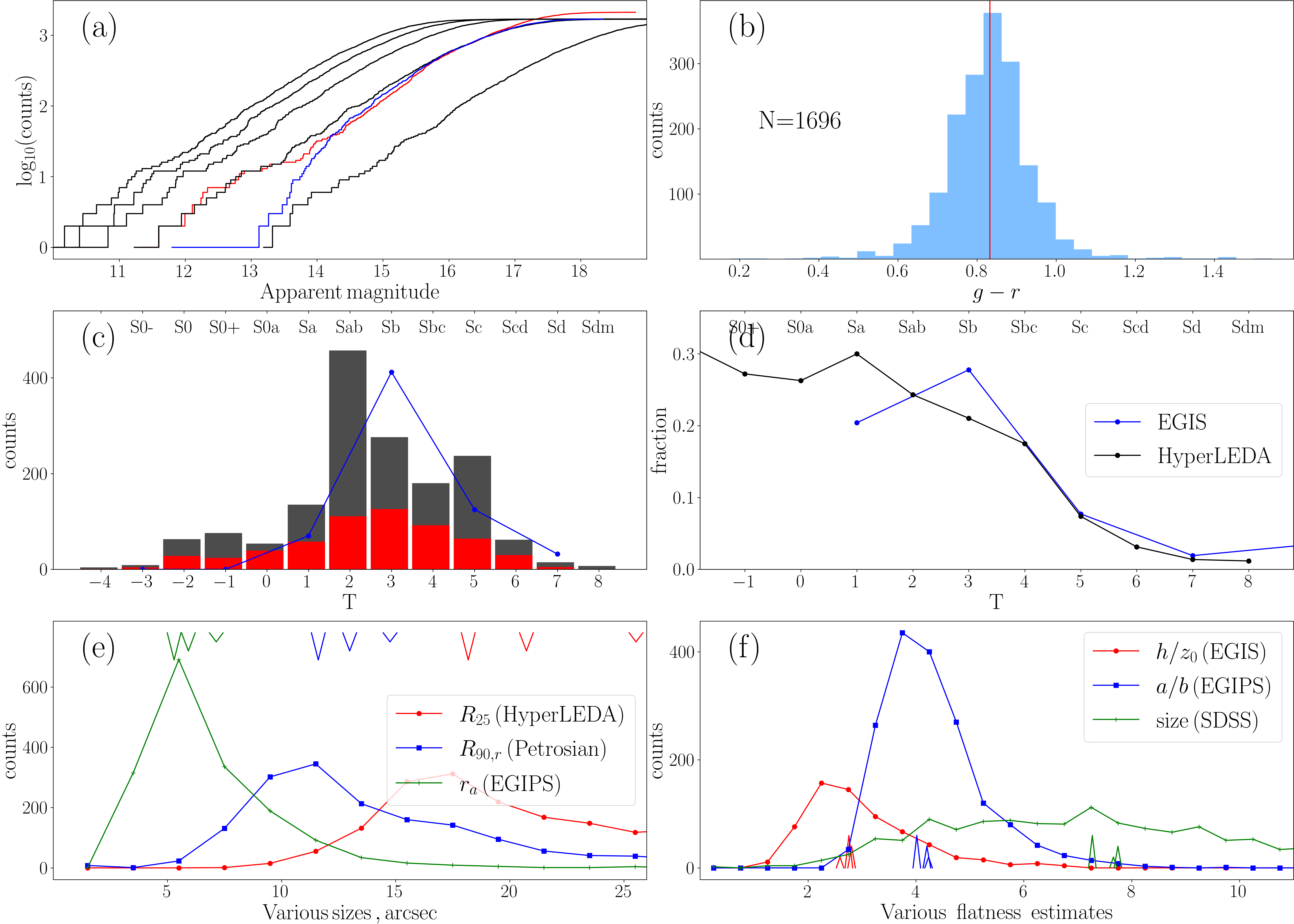}
\caption{The observational properties of the constructed sample discussed in Sec.~\ref{sec:samplepprops}. (a) Cumulative distribution of apparent magnitude, black lines from left to right correspond to the $z,i,r, g,$ and $u$ bands, respectively. The red line corresponds to the $B$ band from HyperLEDA, the blue one is for \texttt{magauto} in the EGIPS $g$ band. (b) Colour distribution $<g-r>$ from SDSS DR16, the line shows the median value $0.85~\mathrm{mag}$. (c) Distributions of morphological types $\rm{T}$ from HyperLEDA, the grey and red histograms are for the entire sample and the $S_{d<150}$ subsample, respectively. The blue line shows the types of the EGIS subsample determined in \citet{Bizyaev_etal2014} (d) Fraction of morphological types $\rm{T}$ in the sample with relation to EGIS and EGIPS data. (e) Distributions of galaxy sizes; small `triangles' at the top show medians for individual classes, the smallest for ``1'' class, the largest for ``3''. (f) Distributions of various measures of galaxy `flatness',
estimated for corresponding subsamples in our sample. Small `triangles' show medians for individual classes, similarly to (e).}
\label{fig:sampleprops4}
\end{figure*}

\subsection{Sample properties} 
\label{sec:samplepprops}

In total, we have 1925 galaxies with X-structures of different quality. In this subsection, we check the observational properties of these galaxies and their distributions. 
\par
The distribution over the sky plane in equatorial coordinates is presented in Fig.~\ref{fig:skypos}. The distribution is inhomogeneous, with two distinct regions that correspond to the regions studied in three individual surveys included in DESI (see figure~1 in \citealp{Dey2018}). The presented sample covers quite uniformly almost all area inside these regions with no visible dependence of location on the X-structure quality class. We also check that positional angles, determined in EGIS and EGIPS catalogues, are evenly distributed across all galaxies and for all quality classes.
\par
We use several sources to improve and double-check the analysis. We collect various data using HyperLEDA\footnote{http://leda.univ-lyon1.fr/} database \citep{leda,hyperleda}, where for each galaxy we find the closest object in coordinates within a radius of 5~arcsec. We use the same scheme to obtain photometric data from SDSS DR16 \citep{Ahumada2020} using `photoTag' and `SpecPhotoAll' tables. Since SDSS and DESI observed regions are not identical, we cannot find a related object for all galaxies in our sample. As a result, we have the data from SDSS for 1512 galaxies. We also use some parameters estimated in EGIS and EGIPS catalogues. All data sources, used in this and the following sections are listed in Table~\ref{tab:inters} and we also mention in the Table the intersection size with the built sample.
\par
For distance estimations, we use the redshift distance modulus \texttt{modz} parameter from HyperLEDA and \texttt{redshift} from SDSS. Both parameters have similar values and we show the one 
that we were able to retrieve for a larger number of objects, i.e. \texttt{modz}. Histograms of distances for each of the quality classes are given in Fig.~\ref{fig:Mabs}, upper subplot. The median distance is about 200 Mpc. As can be seen from the Figure, distribution peaks for individual quality classes located as expected, i.e. ``1'' and ``3'' class objects are closest and furthest on average, respectively. The galaxies from the first class (with clearly visible X-structures) are almost absent at distances $>300$~Mpc, while the other two classes are distributed up to twice as large distance.
\par
 The cumulative distributions for apparent Petrosian magnitudes for 5 SDSS bands are shown in panel (a) of Fig.~\ref{fig:sampleprops4} (black solid lines, $z,i,r,g,u$ from left to right). Median apparent magnitude in the $g$ band equals 16.3~$\mathrm{mag}$ with standard deviation is equal to  1.3~$\mathrm{mag}$. In the same panel, we also show the apparent total $B$ magnitude from HyperLEDA (\texttt{bt} parameter; red curve) and the estimated total apparent magnitude in the $g$ band from EGIPS (blue curve). We can see that the latter tends to show fainter galaxies, but for $m>14\, \mathrm{mag}$ photometry from two $g$ band sources and one $B$ band source are close two each other, as expected. Thus with enough confidence, we can use any of the sources for absolute magnitude estimation. We then use data from SDSS and correct all magnitudes for the extinction in the Milky Way obtained in the SDSS table from the corresponding parameter~\texttt{extinction\_\textit{band}} for each individual \texttt{\textit{band}} under consideration. The median extinction correction in the $r$ band is 0.07~$\mathrm{mag}$ and can reach up to 0.6~$\mathrm{mag}$. The distance--absolute magnitude relation is presented in Fig.~\ref{fig:Mabs} for the $g$ band and has a traditional shape. The right subplot demonstrates the distribution of $M$ for the whole sample and individual subclasses. The mean value of the magnitude $-20.2\, \mathrm{mag}$ is close to that found for EGIS galaxies in the $r$ band (see figure~10 in \citealp{Bizyaev_etal2014}). 
 We see that galaxies of different classes have similar distributions, with galaxies of class ``1'' only slightly brighter than objects of the other two classes (the difference between peaks of distributions of classes ``1'' and ``2'' and classes ``2'' and ``3''  is $\approx 0.1\, \mathrm{mag}$).
 \par
As shown in Fig.~\ref{fig:Mabs}, it is reasonable to analyse additional subsamples of galaxies, namely a subsample of galaxies located closer than 150~Mpc, with 654 galaxies, which includes both faint and luminous objects, and a subsample of galaxies brighter than $-19.5~\mathrm{mag}$, including 1223 galaxies, which correctly represents the distribution by distances. 
For all the results obtained, we check whether they remain the same for these two subsamples, hereinafter referred to as $S_{d<150}$ and $S_{M<-19.5}$, respectively, and mention only those cases when something has changed.
\par
The $<g-r>$ colour distribution for the whole sample, calculated using reddening-corrected values, is shown in Fig.~\ref{fig:sampleprops4} panel (b). All individual classes demonstrate the same symmetrical distribution with a peak close to $0.85~\mathrm{mag}$ (they are not shown). Additionally, using \citet{2006A&A...460..339J} transformations between colour systems, we estimate that the mean $<B-V>$ colour for galaxies is equal to $\approx 1.0~\mathrm{mag}$. With the account for internal extinction in galaxies, which can reach $0.1-0.2~\mathrm{mag}$ in $<g-r>$ \citep{Masters2010}, these colours corresponds to Sa-Sb morphological types and show good agreement with the distribution over the types presented in Fig.~\ref{fig:sampleprops4}, panel (c).
\par
Fig.~\ref{fig:sampleprops4}, panel (c) demonstrates the distribution over the morphological types in numerical codes according to \cite{VaucouleursRC2}. We found that there is no difference in the distributions of types for individual classes and we do not depict them. Compared to figure~17 in~\cite{Yoshino_Yamauchi2015}, where Sc types predominated, our sample shows a slightly different trend. 
The histogram shown in Fig.~\ref{fig:sampleprops4} is, on the whole, significantly shifted towards later types, but most of the galaxies in the sample belong to the Sab-Sb types. 
Thus, our distribution is more consistent with that shown in figure~7 of~\cite{Erwin_Debattista2017}, where the B/PS bulges were found to be distributed nearly equally for all classes before Sbc, with a significant decrease in their number after that. 
A similar drop, starting with the Sb-Sc types, also occurs for the galaxies in our sample, as can be seen from panel (c) in Fig.~\ref{fig:sampleprops4}. 
It is also interesting to note that galaxies from the $S_{d<150}$ subsample exhibit a smoother distribution of types, with the same peak for Sab-Sb types. 
In Fig.~\ref{fig:sampleprops4}, we also plot the distribution of types based on the EGIS classification (blue dots). 
In EGIS, authors used an automatic algorithm, described in \citet{Kautsch2006,Kautsch2009}, to derive the morphological parameters from the concentration index and the ellipticity of the galaxies, which are uniquely related to the $B/T$ ratio (see discussion in Sec.~\ref{sec:BD}). 
Their derived types demonstrate a similar distribution with a peak at $\mathrm{T}\approx3.5$ and drop in the fraction at Sb-Sc. 
\par
There are various estimates of galaxies sizes available in online databases, and we use several of them for greater reliability. 
We found radii of 25th~$\mathrm{mag}$ isophote using \texttt{logd25} parameter from the HyperLEDA database. We also used 
\texttt{petroR90\_r} from SDSS tables, e.g. radius containing 90\% of Petrosian flux in $r$ band, and \texttt{a\_r} parameter from EGIPS tables, which is the standard deviation of the distribution of light along the major axis of the galaxy estimated by \texttt{SExtractor}~\citep{1996A&AS..117..393B}. 
Despite the difference in meaning, all these sizes demonstrate similar behaviour, shown in Fig.~\ref{fig:sampleprops4}, panel (e).
In each case, the apparent size of galaxies on average notably decrease from class ``1'' to class ``3''. 
Galaxies of the class ``1'' have an average size of the semi-major axis of 25th~$\mathrm{mag}$ isophote $\approx 26$~arcsec in the $r$ band, which is 1.5 times larger than for the third class. 
We also found that all three classes contribute to the tail with the largest galaxies, which is again can be expected from the behaviour of distances distribution up to 100 Mpc. 
Note that the peak for the \texttt{a\_r} parameter is at the same location as for the entire EGIPS dataset, as shown in figure~6 in \citet{2022MNRAS.tmp..256M}.
\par
We check major to minor axis ratios in the $r$ band using \texttt{a} and \texttt{b} parameter data from EGIPS analysis and present them in Fig.~\ref{fig:sampleprops4}, panel (f). Galaxies of different classes have very similar distributions with a peak close to $a/b\approx4$. This value is slightly lower than the ratio measured for the whole PAN-STARRS data in EGIPS, see figure~6 in \citet{2022MNRAS.tmp..256M}. Fig.~\ref{fig:sampleprops4}, panel (f) also demonstrates the model-dependent ratio of $h/z_0$ for galaxies from EGIS, where $h/z_0$ is the ratio of radial to vertical scales for discs obtained from 1D decomposition in \citet{Bizyaev_etal2014}. Similarly to $a/b$, galaxies of different classes have the same distribution, but again with a smaller ratio than the whole EGIS sample. For EGIS sample $h/z_0\approx 2.5$ (see figure~4 in \citealt{Bizyaev_etal2014}), however, it should be noted that the presented distribution is limited by the $z_0$ size and for the whole EGIS sample the average ratio is even higher. 
We additionally verified that both the distributions remain the same for the $g$ band. Thus, we can conclude that galaxies with B/PS bulges tend to be systematically thicker than the other edge-on galaxies, but a careful estimation of the size effect is needed in future investigations. In addition, we can note from Fig.~\ref{fig:sampleprops4} panel (f), that the \texttt{size} parameter from SDSS data, calculated using second moment of the intensity, and which is a good measure of the ellipticity \citep{2011AJ....141..189V}, also shows small variation in value between the different classes.
\par
\par
The question of special interest is how common the B/PS bulges are in the local Universe. Some works are entierly dedicated to the frequency estimation of such structures, e.g. \citet{Yoshino_Yamauchi2015,Kruk_etal2019}. 
However, it is difficult to directly compare the obtained fractions between these works, because the samples are formed in different ways, and we need to take into account various kinds of biases.
Thus, in \citet{Erwin_Debattista2017} and \citet{Kruk_etal2019}, the fraction was found to be equal 40\%-50\%, but the studied sample was prepared using only the barred galaxies with moderate inclinations. A greater fraction, 62\% (13 from 21 galaxies), were found in stellar kinematics study of~\cite{Gadotti_etal2020}. In \citet{Lutticke_etal2000}, for 734 galaxies classified as ``edge-on'' with $i~>~60-70 \deg$, authors found that 4.1\%, 15.7\%, and 25.2\% of galaxies have B/PS bulges of different quality, but their criteria are isophote-based, so comparison with their results is not easy. In \citet{Yoshino_Yamauchi2015}, where the decomposition procedure was applied, authors found 292 B/PS bulges in a sample of 1329 edge-on galaxies ($i$-band), 
thus $\approx 22\%$ in total (see Sec.~\ref{sec:validation}). 
\citet{Savchenko_etal2017} identified B/PS bulges with only the brightest, clearly distinguishable X-structures and found 151 candidates from 2021 edge-on galaxies of the EGIS catalogue, so the fraction of galaxies with B/PS bulges was 7.5\%.
Here we found 629 galaxies with B/PS bulges out of a total number of 5733 objects in EGIS during the training sample creation process and 1695 candidates out of 13048 galaxies from EGIPS after voting evaluation, which results in 11\% and 13\%, respectively. These fractions are substantially lower than numbers from other studies. 
The first possible reason for this discrepancy is that we focus only on galaxies in which the X-structure is well distinguishable, while other works usually use less strict criteria, often based on the isophote analysis\footnote{This may also be the reason why we have a lower fraction of the ``3'' class compared to other works, where authors also split data into 3 classes.}. The only work with which we can compare the presented results directly is \citet{Savchenko_etal2017}, but we obtained higher values than these authors. 
\par 
Naturally, bright X-structures should be observed when the bar is seen side-on, that is, when its major axis is perpendicular to the line of sight. The smaller the angle between the bar axis and the line of sight, the less distinguishable are X-structures. For model galaxies and their X-structures in~\cite{Smirnov_Savchenko2020}, we found that X-structures are clearly visible in the residue image up to the bar viewing angle of 45 $\deg$. At 30 $\deg$ and less, the residue images become substantially less informative and lose a characteristic X-shape, although it can still be identified if the resolution is good. Therefore, we assume that we miss about a half of B/PS bulges within our approach.
\par
In addition, as shown by the confusion matrix in Fig.~\ref{fig:trainbyepoch}, the trained ANN can miss some part of the suitable galaxies, but we expect that there will be few of them, and this will not have a significant impact on the final result.
It is also interesting to note that we have evidence of bias related to galaxy masses, as was found by \citet{Erwin_Debattista2017} and confirmed by \citet{Kruk_etal2019} (we also detected it, see Sec.~\ref{sec:stellarmasses}), which also makes direct comparison of numbers difficult.
\par
To conclude this subsection, we want to address a question about some model-dependent parameters of decomposition. We detect statistical differences between the general sample and galaxies with B/PS bulges. Thus, latter demonstrates bigger by 0.5~arcsec vertical scale $z_0$ and larger by $\approx 0.1$ ratio $z_0/h$, larger disc face-on central surface brightness $S_0$ (difference on average $0.9~\mathrm{mag}$) according to 1D decomposition performed in \citet{Bizyaev_etal2014}; larger image smoothness parameter and brighter absolute rest-frame bulge magnitude according to decomposition for galaxies in SDSS DR7 done in \citet{sdss2011}; larger up to $0.5~\mathrm{mag}/\mathrm{arcsec}^2$ disc central intensity and bulge effective intensity according to \citet{Yoshino_Yamauchi2015} decomposition. 
In all these cases, regardless of the implemented decomposition scheme, we also see that galaxies with visible B/PS are generally brighter and, on average, have larger $B/D$.
Probably, some of the differences found are not statistically significant or have little statistical significance, and this issue is important and requires additional research.
To get an idea of how cumbersome this problem can be, we refer the reader to the discussion in \citet{Bizyaev_etal2014} and in \citet{Mosenkov2015}, where the authors compare 1D (fitting the photometric profiles along the major and minor axes), 2D (fitting the whole information from the galaxy image) and 3D (an additional dust extinction model is adopted) decompositions for edge-on galaxies.
As the initial approach to future investigation, in this work we address only the question about $B/D$ values in Sec.~\ref{sec:BD}.

 


\begin{table*}
    \caption{The size of the intersection with the data we are comparing with. The first column shows acronym for the data sample, the next are the reference, sample size, and intersection with our data. The last two columns show whether the sample contains available photometry and decomposition.}
    \begin{tabular}{c|l|c|c|c|c}
    \hline
         Name & Link & Size & Intersection & Photometry & Decomposition \\
         \hline
         EGIS & \citet{Bizyaev_etal2014} & 5,747 & 629 & + & + \\ 
         EGIPS & \citet{2022MNRAS.tmp..256M} & 16,551 & 1695 & + & - \\ 
         Simard et al. (2011) & \citet{sdss2011} & $\approx 10^6$ ($\approx 30,000$ edge-on) & 327 & + & + \\ 
         Zoo DECaLS & \citet{zoodeca} & 314,000 & 588 & - & - \\ 
         HyperLEDA & \citet{hyperleda} & $\approx 10^7$ & 1925 & + & - \\ 
         SDSS DR16 & \citet{Ahumada2020} & $\approx 2.6\times10^6$ & 1512 & + & - \\ 
         Kauffmann et al. (2003) & \citet{Kauffmann2003} & 122,808 & 483 & - & - \\ 
         Yoshino et al. (2003) & \citet{Yoshino_Yamauchi2015} & 1329 edge-on (292 B/PS) & 198 & + & + \\ 
         \hline
         \end{tabular}
    \label{tab:inters}
\end{table*}

\subsection{Sample validation}
\label{sec:validation}

In \citet{Yoshino_Yamauchi2015} authors also divide B/PS bulges in edge-on galaxies from SDSS DR7 into three quality subcategories, specifically called bx+, bx, and bx-. Galaxies from the first category have strong B/PS bulges with the X-shape clearly visible on both sides. For B/PS bulges from the second category the boxy shape is obvious, but the X-shape is less clear. The shape of B/PS bulges marked as bx- is somewhat rectangular, with no traces of the boxy- or X-shape. These criteria are not exactly equal to what we mark as ``1'', ``2'', and ``3'' quality categories, but we still can compare them to check and validate our voting results. There are 198 galaxies in common with \citet{Yoshino_Yamauchi2015}, 153 of them are marked as having B/P structures in their work. The remaining 45 galaxies are mostly of the second or third quality class, but there are also several examples of class ``1'' galaxies with obvious X-structure, so it seems strange that they are not recognised in \citet{Yoshino_Yamauchi2015}. 
The confusion matrix of our classes versus their classes is shown in Fig.~\ref{fig:trainbyepoch}, lower subplot. The agreement is moderately good as seen in the Figure. Thus, the largest numbers in rows are in diagonal cells, with the exception for our first quality class, where the bx number is higher, which, as already mentioned, may be the result of a slightly different methodology or differences in images from SDSS and DESI. We specifically checked the extreme cases, e.g. the galaxy marked as bx+ in \citet{Yoshino_Yamauchi2015} and as a third quality class in our voting. 
It is PGC~2283961, which exhibits a faint boxy bulge even in the DESI residual image, and therefore it should not be considered as a galaxy with a clear X-structure. Examination of three galaxies with marked bx- and ``1'' quality class, namely IC~4020, PGC~091064, and PGC~091593, in all cases revealed clear B/PS bulges. 
\par 
It is also interesting to note the following. The fraction of bx- galaxies in \citet{Yoshino_Yamauchi2015} sample is the largest one. Therefore, one can expect that the intersection of our and their samples should mostly contain the galaxies corresponding to their bx- type. However, this is not true and the intersection is mostly constituted by galaxies of their bx type. This can indirectly indicate that we missed some B/PS bulges with a weak X-shape in our sample (e.g. bx- galaxies). 
\begin{figure*}
\includegraphics[width=1.95\columnwidth]{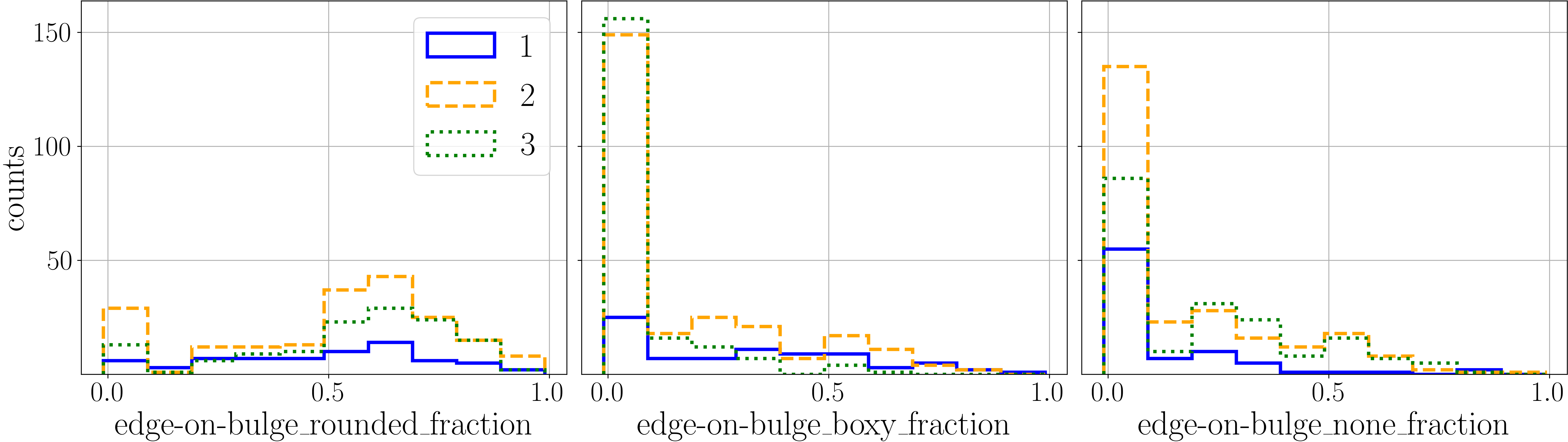}
\caption{Histograms for different bulge vote options from DECaLS Zoo \citep{zoodeca} for galaxies from our sample. Each bin shows the fraction of votes for this options, the options are mutually exclusive. The blue, orange, and green lines are for classes ``1'',``2'', and ``3'', respectively.}
\label{fig:decazoo}
\end{figure*}

\begin{figure}
\includegraphics[width=0.95\columnwidth]{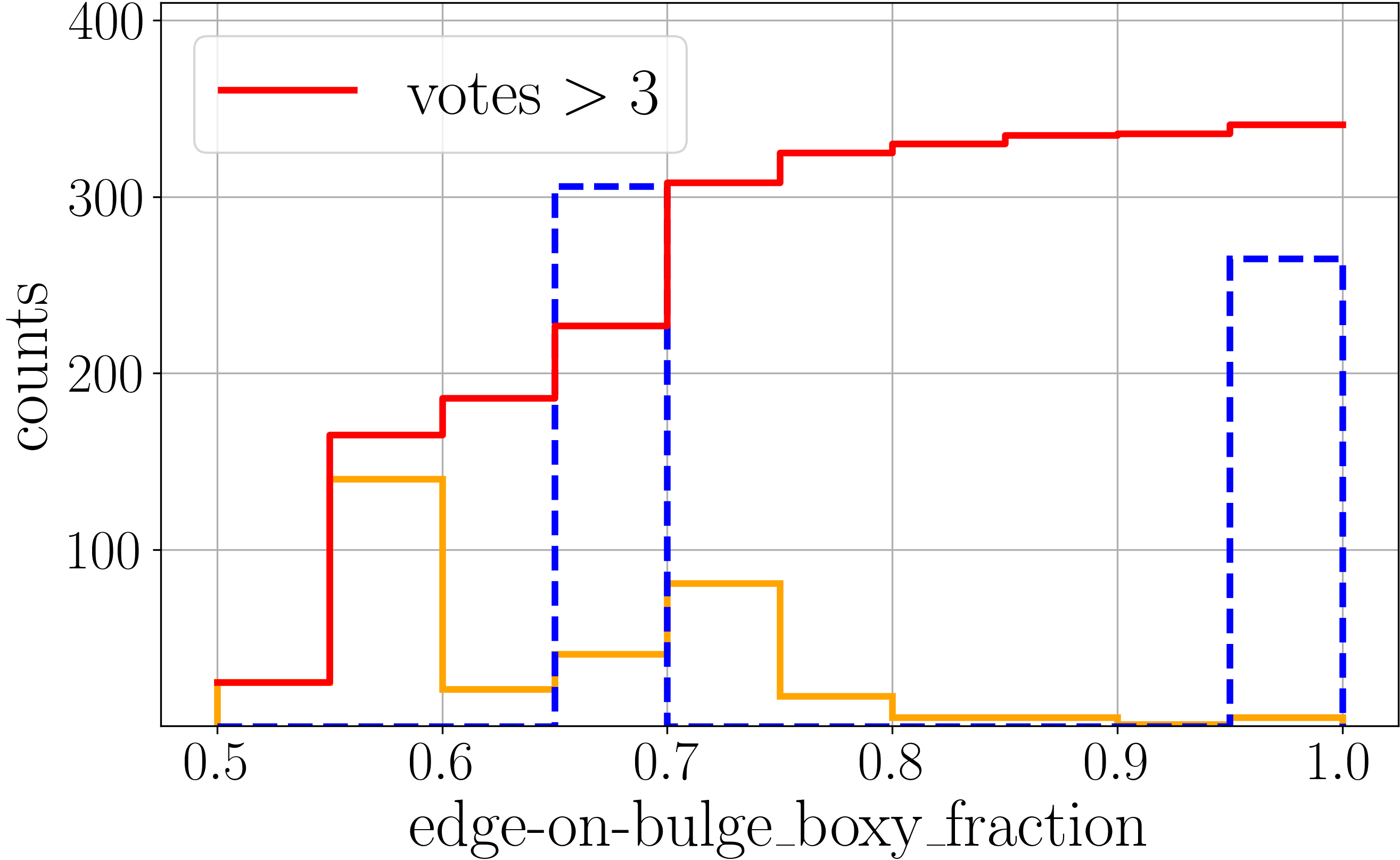}
\caption{Total number of galaxies with `edge-on-bulge\_boxy\_fraction' > 0.5 in the DECaLS Zoo sample. Solid orange bars show those with more than 3 votes per galaxy, dashed blue bars indicate fewer. The red line represents the overall cumulative distribution of the orange bars.}
\label{fig:zooboxy}
\end{figure}

It is interesting to test the sample we created with votes from the Galaxy Zoo project, especially since the recently published version \citep{zoodeca} used images from the DECaLS database, which is part of the DESI Legacy. The authors collected information from volunteers on the properties of 314,000 galaxies, including the question of the shape of the bulge, which is of particular interest here. There are two volunteer catalogues available with approximately 100000 and 260000 galaxies. The difference is that in the latter, called GZD-5, the authors use slightly different decision trees, and most of the objects have much fewer votes (the average value is about 5 votes per galaxy), which, in our opinion, is still acceptable. For consistency and to avoid ambiguity, we use GZD-5 to find the intersection with our sample by coordinates within 3~arcsec, which gives 588 galaxies.
\par
We first ran a simple test to see if all of our galaxies were indeed edge-on. To do this, we looked at the answer `disc-edge-on\_yes\_fraction', which shows the fraction of all votes for a given galaxy that it is edge-on. We can confirm that the vast majority of galaxies in the intersection demonstrate $i \approx 90~\deg$ with very few exceptions, which could be just human errors or small, hard-to-see objects. If a volunteer marked the galaxy as edge-on, then the question was asked about the shape of the bulge (see \citealt{zoodeca} for details and figure~B1 therein). The distribution of answers for each option, which are specifically `boxy bulge', `rounded bulge', and `no bulge', is shown in Fig.~\ref{fig:decazoo}. Each panel shows the distribution of the fraction of `yes' answers to each binary question.
It should be noted that the same galaxy may have votes for mutually exclusive options, e.g. for a square and rounded bulge. The plot for `edge-on-bulge\_boxy\_fraction' shows the percentage of approved B/PS bulge votes for edge-on galaxies, which is exactly what we are trying to find in this paper and therefore the matter of a particular interest to us. Contrary to our expectations, there are practically no approved square-shaped bulges. It is worth noting the difference between the distributions of each quality class and the fact that class ``1'' galaxies show a much steeper decline. In total, we have only 49 galaxies with indicated boxy-shaped bulges, if we consider those for which the share of votes cast is more than 0.5. The reason why we hardly see B/PS votes in the Zoo data is discussed at the end of the subsection. Interestingly, another option of the `edge-on-bulge \_rounded\_fraction' of the same question shows a completely different picture, with most volunteers seeing bulges with an X-structure as round objects. The likelihood of ``flatness'' of bulges tends to decrease from ``3'' to ``1'' classes, and the distribution for class ``1'' galaxies is much more homogeneous, suggesting ambiguity among participants. The last option `edge-on-bulge\_none\_fraction' shown in Fig.~\ref{fig:decazoo} indicates that volunteers rarely flag galaxies in question as having no bulge or a small one.
\par
\par
In order to address an issue of the unexpected lack of votes for boxy bulges at the intersection, we decided to check how many of such objects can be found in total in \cite{zoodeca} data (also check their figure~10). We checked both volunteers' catalogues separately and calculated how many galaxies have at least `edge-on-bulge\_boxy\_fraction' > 0.5. In addition, we excluded cases with one vote and isolated cases with 3 or fewer votes (in particular, `disc-edge-on\_yes votes' > 3, a question about the shape of a bulge, asked only for edge-on galaxies) from those for which the number of votes allows making more informed conclusions. The histogram of `edge-on-bulge\_boxy\_fraction' for these cases and the cumulative distribution for the reliable number of boxy bulges are presented in Fig.~\ref{fig:zooboxy} for GZD-5\footnote{The second catalogue shows similar numbers, and we decided not to show it.}. We see that cases with a small number of votes differ from the more reliable ones with > 3 votes. For the latter, even if we include cases with votes fraction starting from 0.5, we can barely collect 350 galaxies with boxy bulges (red curve). 
Moreover, for more reasonable conditions, such as, for example, a fraction $>0.75$, we only can obtain a few dozen. 
Since we also notice a significant amount of votes for questions such as  `has-spiral-arms\_yes\_fraction' or `spiral-winding\_tight\_fraction', which clearly should not be presented for verified edge-on galaxies, we expect that cases with a small number of votes are very doubtful and the number of reliable B/PS bulges is less than what is found in them. Even if we also sum up all cases including those with 3 or fewer votes, in total we will see less than a thousand examples. 
To find such a number from a sample of 314000 galaxies with approximately 20-40 thousands placed edge-on seems to be much less what we can expect from an estimate of the B/PS fraction in \citet{Yoshino_Yamauchi2015,Lutticke_etal2000,Erwin_Debattista2017,Kruk_etal2019}. This extreme rarity of boxy bulges in \cite{zoodeca} should be the reason why we obtain such results when comparing Galaxy Zoo DECaLS votes with our sample. This can also be implicitly confirmed by the fact that if galaxy marked as not edge-on, volunteers can vote for the presence of a bar, and indeed they see a bar feature in a noticeable number of definitely edge-on cases, which we can speculate to be a sign of boxy bulge instead. Overall, such a small number of B/PS found even for large samples like Galaxy Zoo clearly emphasise the value of our work and shows how difficult the task is due to the already mentioned fact that the B/PS feature can be faint, and why the usage of residual images instead of regular ones is important.
\par
To conclude this subsection about sample validation, we want to note that the distributions in Sec.~\ref{sec:samplepprops} are in line with expectations, as well as with all consequent analysis in the next Section, especially in Sec.~\ref{sec:BD}, \ref{sec:angles}, and \ref{sec:stellarmasses}, and these results can be viewed as a kind of check too.

\subsection{Interesting objects}
\label{sec:interestingobj}

\begin{figure*}
\includegraphics[width=0.43\textwidth]{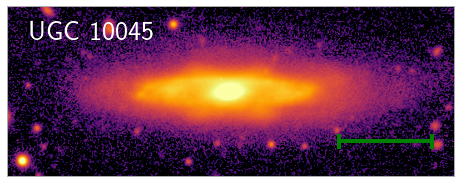}
\includegraphics[width=0.43\textwidth]{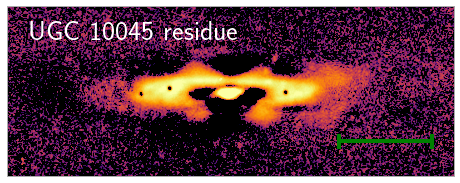}\\
\includegraphics[width=0.43\textwidth]{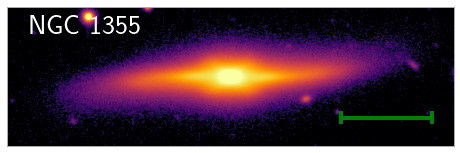}
\includegraphics[width=0.43\textwidth]{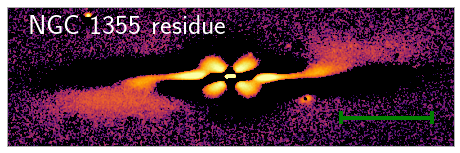}\\
\includegraphics[width=0.43\textwidth]{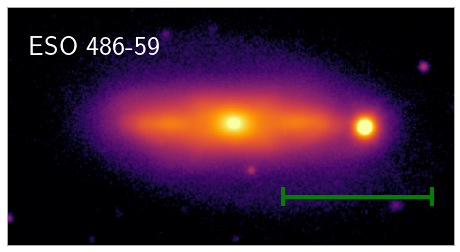}
\includegraphics[width=0.43\textwidth]{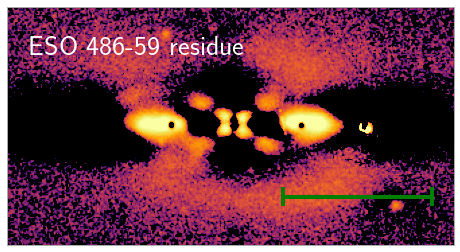}\\
\includegraphics[width=0.43\textwidth]{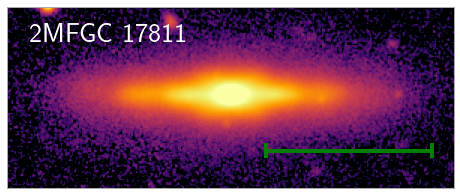}
\includegraphics[width=0.43\textwidth]{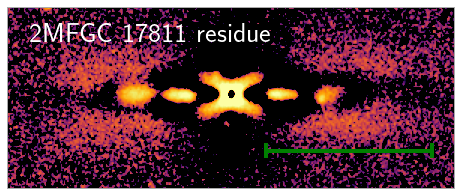}\\
\includegraphics[width=0.43\textwidth]{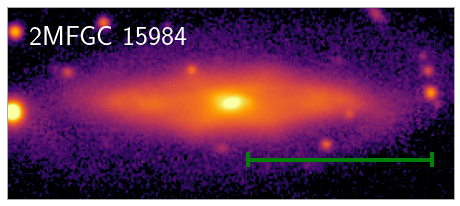}
\includegraphics[width=0.43\textwidth]{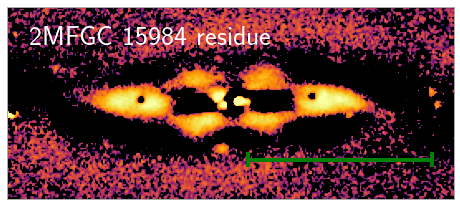}\\
\includegraphics[width=0.43\textwidth]{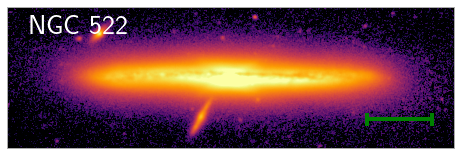}
\includegraphics[width=0.43\textwidth]{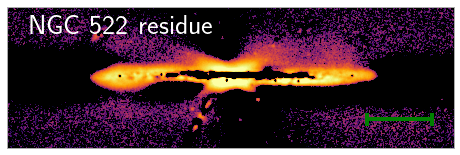}\\
\caption{A short list of interesting objects from Sec.~\ref{sec:interestingobj}: original images in the $r$ band (\textit{left column}) from DESI Legacy and residue images~(\textit{right column}). The green bar marks the scale of 30 arcsec.}
\label{fig:objects}
\end{figure*}

In this subsection, we provide a short list of some interesting objects, which was compiled based on the participants notes during the voting step. This list is not uniform in the sense that some objects have quite intriguing features or geometry, while in some cases there are just very strong bar and X-structures. All but one (NGC~522) of the galaxies below have never been studied for B/PS bulges and their properties as far as we know. Thus, it is this work that allowed us to highlight them explicitly. The images and their corresponding residues for all discussed objects are presented in Fig.~\ref{fig:objects}.
\par
\textbf{UGC 10045}: This galaxy is observed nearly edge-on, but has a slight apparent tilt. It shows a bright B/PS bulge, albeit without such sharp X-structure. Its rays manifest themselves in the residue image. The bulge itself is very large in comparison with the entire galaxy, its linear size is about half of the disc. We should also note that the B/PS bulge is attached to the thick ring with almost no gaps in between, which means that a thin (flat) part of the bar have very small linear size (if any) in this galaxy. This phenomenon is especially interesting in light of~\cite{Patsis_2021}'s study of NGC 352 galaxy and works by~\cite{Erwin_Debattista2017} and \cite{Smirnov_Sotnikova2018}, where it was found that the typical size of the B/PS bulge should be about half or less than that of the bar. Interestingly, there is also an inner bright feature that resembles a nuclear disc or a small inner bulge. This separate component has a clear trace in the residue image.
\par
\textbf{NGC 1355}: Here one can point to bright and sharp X-structures. In addition, although the B/PS bulge looks symmetric with no sign of bending or buckling, the entire disc is warped and has an ``S'' shape. From our subjective point of view, this is rather an exemplary B/PS bulge, and, therefore, it is very interesting, that NGC 1355 was not specifically studied in any of the following works:~\cite{Ciambur_Graham2016,Laurikainen_Salo2016,Savchenko_etal2017}, where a comprehensive analysis of B/PS bulges of some galaxies was carried out. 
\par
\textbf{ESO 486-59}: The B/PS bulge of this galaxy has a fairly boxy profile and is quite large, which makes it somewhat similar to the case of UGC~10045. But the structure of the residue is much more complicated here. First, there are large `shoulders', which are extensions of the thin part of the bar in the disc plane. Secondly, the inner and outer X-shaped features can be seen in the centre of the residue image. Double X-structures have already been found and discussed in several numerical and observational works~\citep{Ciambur_Graham2016,Smirnov_Sotnikova2018,Parul_etal2020,Ciambur_etal2021}. 
Since ESO 486-59 is not directly observed edge-on and has double X-shaped features that are mostly seen in the residue, it is difficult to say whether such a residue is a direct trace of double X-structures or not. However, this makes this galaxy an interesting target for future research and, in particular, decomposition studies.
\par
\textbf{2MFGC 17811}: Here, the entire B/PS bulge has a square-like profile, but at the same time, an unusual square-like central part is present in the residue. Again, there are large bar `shoulders', after which only a small part of the disc remains. The bar occupies most of this galaxy.
\par
\textbf{2MFGC 15984}:  An X-structure in this galaxy is not as sharp, but manifest itself clearly in the residue image.  Again, the entire galaxy is mostly composed of a bar and a B/PS bulge. There are large `shoulders' that are probably mixed with a thick ring. In the central region, a bright central component can be identified, resembling a disc or lens. It is possible that this component has a flat structure without a thick vertical part, since it is noticeably stretched in the disc plane.
\par
\textbf{NGC 522} 
Although this galaxy appears to have a fairly usual B/PS bulge and has already been studied in~\cite{Laurikainen_etal2014}, its residue is interesting. It shows clear asymmetry with respect to the major axis of the galaxy. Asymmetry can be an indicator of the bar buckling, which is very rare in the local Universe~\citep{Erwin_Debattista2017,2021ApJ...909..125X}. Although the dust lane here is also slightly bent towards the upper side of the galaxy, this factor can lead to an unusual residue structure. We provide the corresponding original image and the residue as the suspicious example of a buckled galaxy. 
\par

\section{Boxy/peanut-shaped bulges analysis}
\label{sec:prelim_analysis}








\subsection{Bulge-to-total ratios}
\label{sec:BD}

\begin{figure*}
\includegraphics[width=1.95\columnwidth]{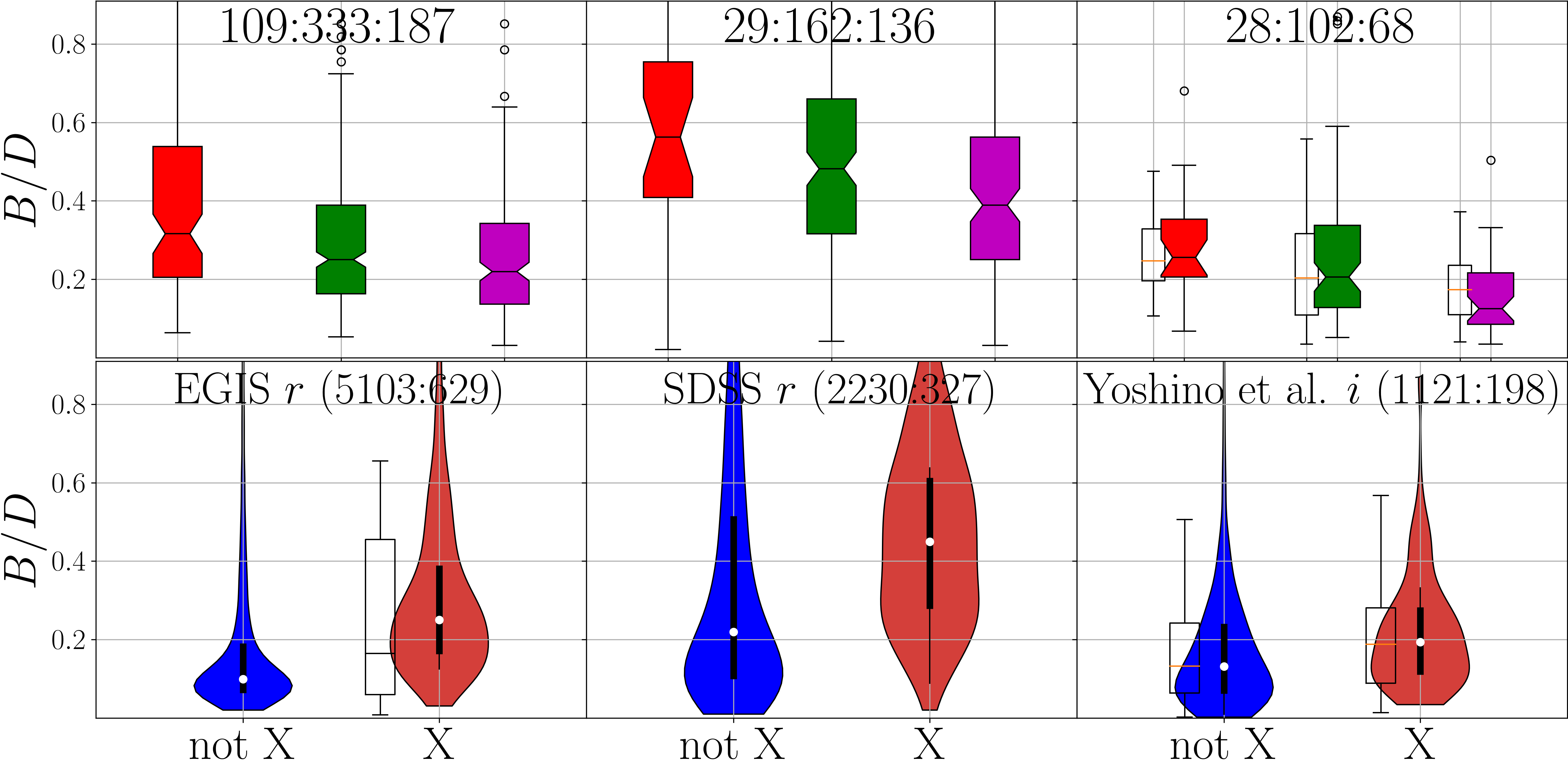}
\caption{Distributions of the  $B/D$ ratio for different decompositions. Each distribution is represented as a so-called boxplot or violinplot. In each case, the median, 25th and 75th percentiles are marked. From left to right, the works of \citet{Bizyaev_etal2014}, \citet{sdss2011}, and \citet{Yoshino_Yamauchi2015} are used. The bottom row shows the distributions of the  $B/D$ ratio for galaxies without (blue) and with (red) an X-structure. The top row shows the distributions for individual quality classes. The band used and the number of galaxies in each case are written in the top of each subplot. The boxplot in the lower left image is for data from \citet{Savchenko_etal2017}, and in the right column for classes from \citet{Yoshino_Yamauchi2015} classification (see Sec.~\ref{sec:BD} for details). }
\label{fig:allbd}
\end{figure*}

In their study of NGC~4565 composite bulge,~\cite{Kormendy_2010} found that a $B/D$ ratio, where $B$ and $D$ are the luminosities of the bulge and disc, respectively, which is usually interpreted in the context of galaxy merging or the secular evolution of galaxies, can be greatly overestimated for edge-on galaxies possessing B/PS bulges. In particular, the overestimation occurs if the central component is described by only one density profile. In such a case, the resulting luminosity of the central component is roughly the sum of the luminosities of all different physical components that can possibly reside in the central region of a galaxy. Among such components can be a lens, a barlens, a B/PS bulge, a flattened pseudo-bulge, a classical bulge, presumably formed during merging of galaxies, and some others. Various tests~\citep{Gadotti_2008, Mendez-Abreu_etal2014, Laurikainen_etal2018} have shown that taking into account the structure of the bar and barlens in the photometric decomposition procedure can strongly reduce the values of $B/T$ ratio (up to two times in~\citealt{Gadotti_2008}). Therefore, formally obtained high values of $B/T$ ratio can be due to the rough decomposition and the merger events may be less important for galaxies with B/PS bulges.  
The question arises as to what $B/T$ ratios are typical for the galaxies in our sample. Since our galaxies should possess the B/PS bulges, we can expect that $B/T$ for galaxies in our sample will also be quite high on average, as in the case of NGC~4565, if we work within the framework of the standard $B+D$ decomposition. 
\par
To verify this, we first compare our results with the decomposition done for EGIS sample in \cite{Bizyaev_etal2014}. There are 629 objects at the intersection of their and our samples. The  distribution of the bulge-to-disc ratio $B/D$ in the $r$ band for these galaxies is presented in Fig.~\ref{fig:allbd} on the top and bottom left plots. The top plot shows distributions boxplots for different quality classes of our sample, with the number of galaxies of each class indicated in the title. We see a noticeable increase in the average of $B/D$ from  class ``3'' to class ``1'', with the median value of about 0.3 for the entire sample. This is again a sign of the consistency of quality classes estimates. 
The bottom plot shows the so-called `violinplot' of $B/D$ distributions for galaxies with B/PS bulges compared to the rest of the EGIS sample. It is easy to notice how different these distributions are. The average value of $B/D$ for galaxies not included in the sample is approximately 0.1. This is almost three times less than the average $B/D$ for those with the an X-shape feature. Comparison of quartiles show that 75\% of galaxies with an X-structure demonstrate strictly higher $B/D$ value than 75\% of galaxies without it. We especially checked the situation with the other bands ($g$ and $i$) and found that there are trends similar to those described above. 
\par
\par
The second source we will compare with is \cite{sdss2011}. The authors have decomposed the $g$ and $r$ images of over a million galaxies from SDSS DR7 catalogue. We found the intersection of our and their samples, identifying the closest SDSS DR7 object within 7~arcsec. In addition to this, we intersected the resulting subsample with the EGIPS data used in Sec.~\ref{sec:sample_building} to find candidates. In total, there are 2566 galaxies in common, of which 327 have an X-structure, which is not that much, but still sufficient for our purposes. As a simple check, we verified the inclinations from~\cite{sdss2011} for galaxies at the intersection. Almost all galaxies exhibit $i\gtrsim80-85\deg$, which additionally confirms the decomposition of~\cite{sdss2011}, where inclination treated as a free parameter.
\par 
In \cite{sdss2011}, the authors use and compare three different models, including the pure Sersic model and the composite disc + bulge model with the Sersic index $n$ set to 4 (de Vaucouleurs bulge) and without this assumption, i.e. with a free value of $n$. We use the $B/D$ values obtained for the model with free $n$ in the $r$ band. The results are presented in Fig.~\ref{fig:allbd} on the top and bottom middle plots. The distributions demonstrate the same trends as the distributions depicted for EGIS case, but with a significantly higher average $B/D$. For galaxies with identified X-structures, the average value of $B/D$ is close to 0.4, while the rest of the intersection demonstrates a half value. It is important to note that we are comparing galaxies that are simultaneously in the EGIPS and~\cite{sdss2011} samples, because we are more confident in them and for reasons of consistency. If instead we take all 31397 near edge-on galaxies from \cite{sdss2011} with $i>83\deg$ and check their $B/D$ distribution, it will be similar to the blue one on the left bottom plot with an average value of $B/D$ close to 0.1, which does not change the overall picture, although some unknown part of these galaxies should also possess B/PS bulges. 
\par
\citet{Yoshino_Yamauchi2015} also applied the 2D decomposition procedure to their sample of galaxies. The decomposition itself was carried out on the basis of 40 template models with different parameters, and the value of $\chi^2$ was minimized by their own code. The $B/D$ values obtained from the decomposition for 198 galaxies in the intersection are shown in Fig.~\ref{fig:allbd} on the right subplots. As can be seen from the figure, the difference between galaxies with and without X-structures is smaller compared to the results of the decomposition considered above, but still exists. The difference is probably smaller here, because some of our "not X" galaxies are actually disguised as `bx' types from~\citet{Yoshino_Yamauchi2015} (average B/PS bulge in their nomenclature).
The trend between quality classes is the same here as before, where class ``1''  galaxies demonstrate the highest values of $B/D$. We also checked how the $B/D$ values change depending on the classes from classification scheme of~\citet{Yoshino_Yamauchi2015} (see Sec.~\ref{sec:validation} for a detailed comparison). The corresponding distributions are presented in the same plots as empty boxes. One can see that $B/D$ is again larger for galaxies labelled as `bx-',`bx', and `bx+' than for the rest of the sample, which is also evident from figure~17 in \citet{Yoshino_Yamauchi2015}. The trend between the individual classes is also similar, but their `bx-' galaxies, on average, demonstrate higher values than our class ``3'' galaxies.
\par
Thus, we can see that regardless of the decomposition technique, in all three mentioned works~\citep{sdss2011,Bizyaev_etal2014,Yoshino_Yamauchi2015} we observe the same result for $B/D$ ratio. This ratio appears to be larger in galaxies with B/PS bulges and massive X-structures. We want to make a few important notes here. First, we can see from Fig.~\ref{fig:allbd} that the average value of $B/D$ varies drastically between individual works, e.g., in \citet{sdss2011} the values are the highest, while in \citet{Yoshino_Yamauchi2015} they are the lowest. There is no easy way to explain the difference, but we suspect the decomposition in \citet{sdss2011} is less accurate than the others, because it is employed for a much larger sample and because it is the only decomposition where the disc scale $h$ is not strongly correlate with the geometric size of X-structure rays (see Sec.~\ref{sec:xsizes}). Secondly, we do not use the values from \texttt{TRACTOR} models \citep{tractor}, because individual components are difficult to attribute properly for the correct galaxy. Thirdly, at the preparation step, we also tried to use decompositions from S$^{4}$G catalogue in \citet{Salo2015}, but the intersection of our and their samples includes only 14 objects and therefore can hardly be compared to the other decompositions with enough significance. 
\par
Concluding this section, we obtained that galaxies with X-structures have a significantly higher value of $B/D$ compared to galaxies that do not have noticeable X-structures. As stated at the beginning of the section, this can be explained as a consequence of $B+D$ type of the decomposition, because if one distinguishes only one central component, the light coming from the B/PS bulge is mixed with the light coming from either the classical or pseudo-bulge (or both), and all type of bulges contribute to $B$. Thus, the results support the idea that the galaxies in our sample indeed posses B/PS bulges, which is what we expect. An additional support for this statement comes from the results of~\cite{Savchenko_etal2017}. In this work, the bulge and disc components were carefully separated from the light of the X-structure for 22 galaxies with B/PS bulges, which made it possible to more accurately estimate $B/D$ for such cases. The computed distribution of $B/D$ in the $r$ band is presented as overlaid data in the right distribution in lower left corner in Fig.~\ref{fig:allbd}. One can see that for many galaxies in \cite{Savchenko_etal2017} the $B/D$ indeed becomes smaller (by about 0.14) if the X-structure is excluded, but the median value is still twice twice as large as for galaxies without B/PS bulges. The latter means that X-structures constitute only a part of B/PS bulge, and a more accurate account of the entire B/PS bulge is required to reconcile the left and right parts of Fig.~\ref{fig:allbd}. 
\par 
As an additional check, we have roughly estimated the typical $B/D$ values for some model galaxies from~\cite{Smirnov_Sotnikova2018} possessing large B/PS bulges using the $B+D$ decomposition. We found that, depending on the strength of the B/PS bulge, bar viewing angle, and disc inclination angle, the values span the range from about 0.3 to 1.1, with a preferred value of about 0.5 and higher. Such scatter of values is generally consistent with the scatter of values observed in the middle plot of Fig.~\ref{fig:allbd}. 
\par
\subsection{X-structure angles}
\label{sec:angles}


The last question in our Zooniverse voting list concerned specifically X-structure parameters. The participants were requested to mark the ends of rays, see the examples of such marks in Fig.~\ref{fig:quality_examples}. We collect 10755 answers to this question and, on their basis, estimate the opening angles and sizes of X-structures. 
Each participant also had an additional option to mark the galaxy centre. Since the position of the centre can be difficult to pinpoint visually, especially in the residual image, we only consider the coordinates of the center when half of the X-structure is obscured by dust in galaxies that are not strictly edge-on. In our sample, we have labelled such cases by index `p'. They constitute only a small part (around 14\%) of the entire sample. For all other cases, we measure the angle between the diagonals of the rectangle with the vertices at the ends of the rays. The opening angle $\varphi$ of the X-structure is equal to half of $180~\deg$ minus the angle between the diagonals. It should be noted that our method does not allow studying the X-structures with different angles below and above the disc plane (vertically asymmetric bars) and, in general, large error bars are expected to be obtained. However, we hope that, due to the significant sample sizes and averaging of individual measurements, the distribution should be close to real, which seems to be the case as the results below show.
\par
\begin{figure}
\includegraphics[width=0.95\columnwidth]{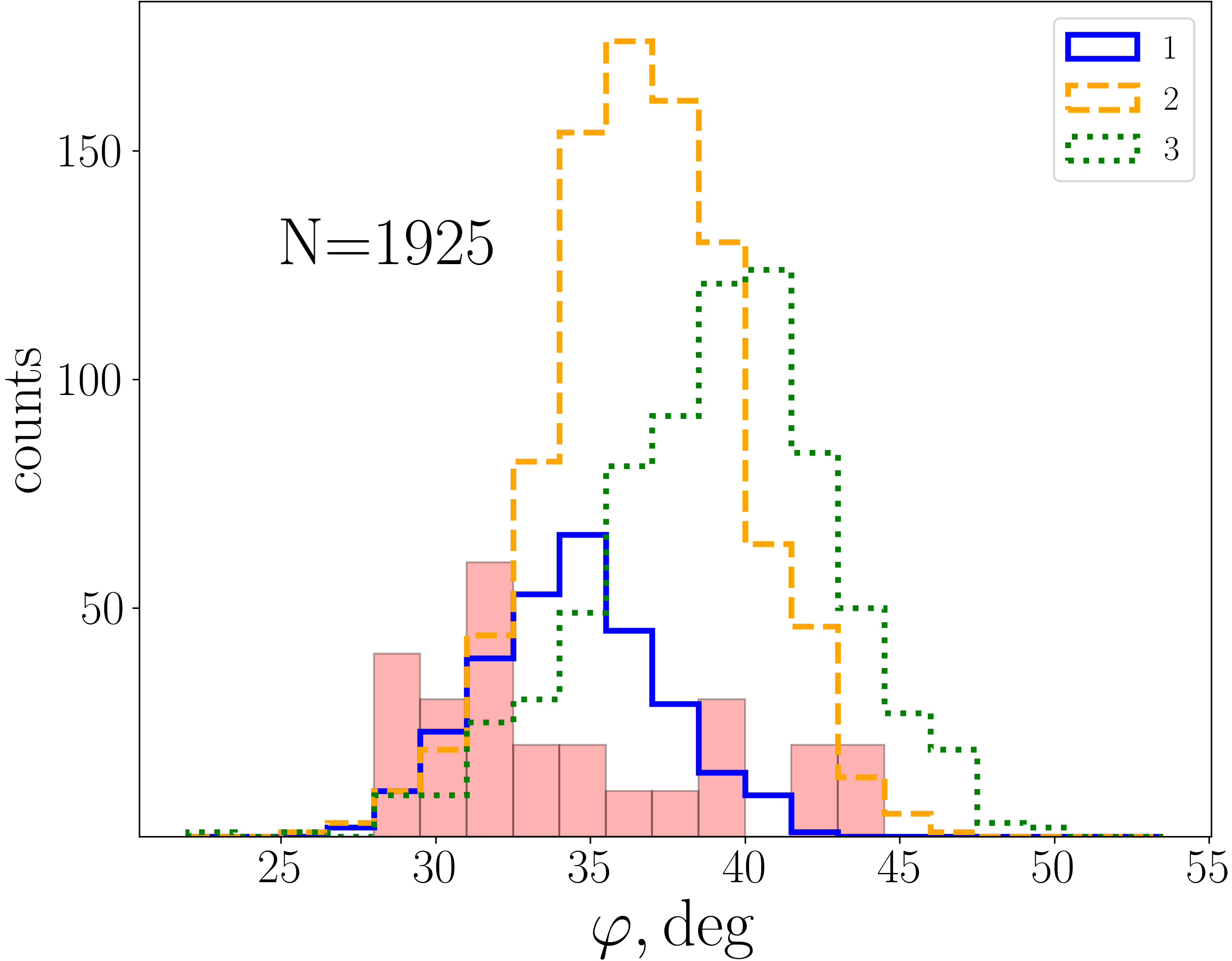}
\caption{The distribution of X-structure opening angle $\varphi$ values for our sample. The blue, orange, and green lines refer to classes ``1'',``2'', and ``3'', respectively. The red bins show the distribution for $\varphi$ from \citet{Smirnov_Savchenko2020} analysis (bin sizes are multiplied by a factor of 10 for better visibility).}
\label{fig:angles}
\end{figure}

 
We averaged the angles over the individual measurements for each galaxy under consideration. 
We checked the uncertainty in the angle estimates, i.e. standard deviation of measurements, and found that its distribution is almost the same for all quality classes, with a mean uncertainty approximately equals $3\deg$.
The resulted distribution of the opening angles $\varphi$ is shown separately for individual classes in Fig.~\ref{fig:angles}.
For comparison, we also depicted the distribution of the opening angles from more careful analysis in~\cite{Smirnov_Savchenko2020} in the same figure. It is easy to see that class ``1'' demonstrates the smallest scatter in the angles, while class ``3'' has the largest scatter on average and also the largest average $\varphi$ value. It is worth noting that almost all angles exceeding $45\deg$ comes from  class ``3'', which consists of the noisiest and faintest objects. 
\par
It is interesting to compare our results with those previously obtained, especially with figure~21 from \cite{Smirnov_Sotnikova2018}. This comparison give us several insights. First, despite the extremely large sample used and the inaccuracy of individual measurements, all measured angles fully confirm the $N$-body predictions from \cite{Smirnov_Sotnikova2018}, where the minimal value of the X-structures opening angle was found to be equal to $25\deg$ in models with a significant contribution of the dark matter within the optical radius of the disc. We note that the projection effects are irrelevant if we compare the lower end of angle values of real and simulated galaxies, since the minimal angle values should be observed for side-on or close to side-on bars~(see figure 22 in~\citealt{Smirnov_Sotnikova2018}).
Secondly, we can see that the estimates for class ``1'' cover almost the same range of values as in~\cite{Smirnov_Savchenko2020}. 
If we compare the mean and median values of the angles from~\cite{Smirnov_Savchenko2020} and from our class ``1'', they turn out to be quite close, namely 34.5 $\deg$ and 33 $\deg$ versus 34.4 $\deg$ and 34.5 $\deg$, respectively, which means that they are consistent with each other. 
We did not find any difference in $\varphi$ distributions for the $S_{d<150}$ and $S_{M<-19.5}$ subsamples, nor did we observe any significant correlations between the angle and collected physical properties of the galaxies in the sample.
\par
Unfortunately, from the presented distribution, little can be said about the physics of real galaxies and their B/PS bulges. The large scatter of values observed for each of the classes in Fig.~\ref{fig:angles} is determined by two independent factors: physical parameters of galaxies and projection effects. By projection effects, we basically mean the dependence of the angle of X-structures on the position of the bar relative to the line of sight~\citep{Savchenko_etal2017,Smirnov_Sotnikova2018,Smirnov_Savchenko2020}. A small inclination of the disc plane also slightly changes the angles, but its effect should be significantly less than the bar rotation (see figure~7 in~\citealt{Smirnov_Savchenko2020}). In the cited works, it was found that the rotation of the bar major axis from a side-on position by about 60 $\deg$ increases the angle of the X-structure by about 10 $\deg$ (see figure~6 in~\citealt{Smirnov_Savchenko2020}). The scatter of values observed in Fig.~\ref{fig:angles} is about the same. But it does not mean that the observed distribution of angles can be fully explained only by the projection effects. First of all, even if the projection effects artificially broaden the distribution, the peak in the distribution is probably associated with galaxies where the major axis of the bar is perpendicular to the line of sight. This assumption is based on the fact that, in this geometric configuration, the B/PS bulge stands out most clearly against the disc background. And the closer the major axis of the bar is to the line of sight, the weaker the B/PS bulge becomes and, therefore, the less likely it is to be detected. The histogram, presented in Fig.~\ref{fig:angles}, has a peak at about 35 $\deg$. In~\cite{Smirnov_Sotnikova2018}, the models that favour such a value of the X-structure angle are models that have either an additional central mass concentration, like a classical bulge (which is quite possible for a real galaxy), or models in which there is no classical bulge, but the disc is initially dynamically warm (the typical value of the Toomre $Q$ parameter is 1.6) or thick~(the typical ratio of the vertical scale of the disc to the  scale in the plane is about 0.1). It is clear that it is impossible to distinguish between these cases by one value of the angle. It is also quite possible that the mentioned model cases do not describe the complete picture. However, it is still interesting that we observe a clear peak in the distribution. Moreover, as shown in~\cite{Parul_etal2020}, a particular value of the angle can be associated with a specific family of regular orbits that create B/PS bulges, and our findings can help one to understand which  families of orbits should actually be present in real galaxies. 


\subsection{Geometric sizes of X-structures}
\label{sec:xsizes}

Another intriguing question is how large the X-structures do we observe in the disc scale and in absolute units. This is important metric to compare with because the size of X-structures correlates with the bar size and tends to increase as the bar grows~\citep{Erwin_Debattista2017,Smirnov_Sotnikova2018}. We estimate the sizes of the rays by measuring the distances from the centre to the points marked by each participant during the voting procedure, and averaging them over all measurements for an  individual galaxy. It is worth noting that the geometric sizes will be even more subjective,  because, unlike angles, they depend on the absolute point positions on the ray, while for angles it does not matter, where exactly the point is located on the ray.
\par
\begin{figure}
\includegraphics[width=0.95\columnwidth]{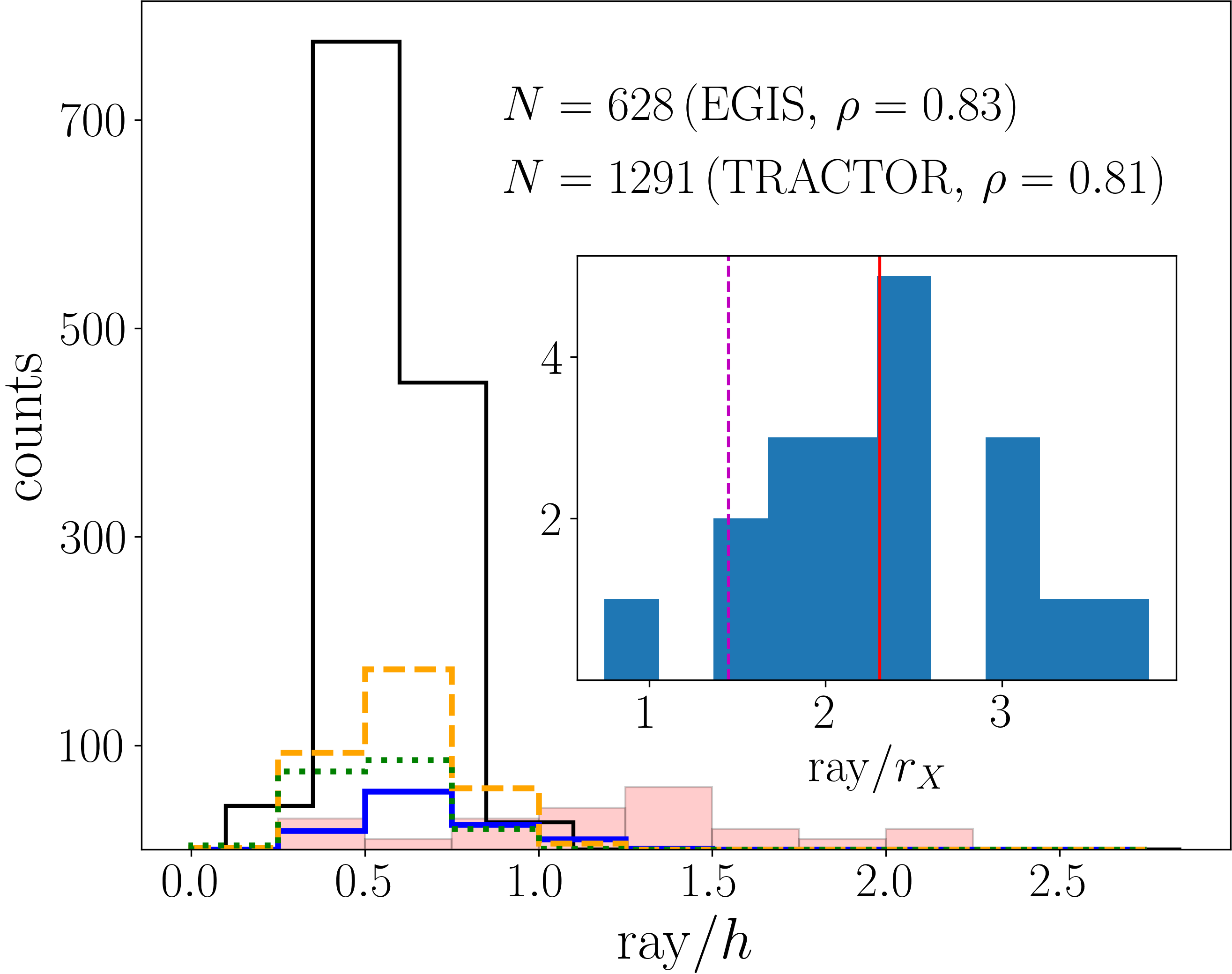}
\caption{Same as in Fig.~\ref{fig:angles}, but for $\mathrm{ray}/h$. The black curve represents this ratio for galaxies with $h$ determined using \texttt{TRACTOR} (see text). The red bins show the same ratio from \citet{Savchenko_etal2017}, multiplied by a factor of 10 for better visibility. The inner plot shows lengths of X-structure rays for galaxies from \citet{Savchenko_etal2017} in scales of $r_\mathrm{X}$, where $r_\mathrm{X}$ is a characteristic scale of the photometric model of the B/PS bulge, which was introduced in~\citet{Smirnov_Savchenko2020}. The solid red vertical line shows the median value for $\mathrm{ray}/r_\mathrm{X}$ for the data from \citet{Savchenko_etal2017}, and the dashed magenta line shows the median value for the same galaxies measured in our sample.}
\label{fig:ray_to_h_EGIS}
\end{figure}

The absolute median value for the rays is close to 3 arcsec. Usually, the size of the X-structure is normalised to the exponential scale $h$ of the disc. Since we do not have the decomposition data available for all galaxies in the sample, we use EGIS $h$ from 1D decomposition in one third of the cases and $h$ calculated from half-light radius found in \texttt{TRACTOR} for the rest where it is possible. As mentioned earlier, the latter models may not be reliable, but since they demonstrate approximately the same distribution (mean $\mathrm{ray}/h$ is 0.56 versus 0.60 for $h$ from EGIS), we decided to use them. The scaled distributions of the ray lengths are shown in Fig.~\ref{fig:ray_to_h_EGIS} for all classes. First, all three classes for decomposition parameters from EGIS show peaks at similar, but different values: the mean value for the class ``3'' is about 0.54, class ``2'' has a mean value of about 0.6, and class ``1'' of about 0.69. The shift towards high values from class ``3" to class ``1'' is quite consistent with our definition of classes, since class ``1'' should consist of galaxies with bright, easily detectable X-structures. 
Naturally, for such X-structures, the ray can be traced to larger distances. We have also found a high correlation between the disc exponential scale and the X-structure ray size (0.83 and 0.81 for $h$ from EGIS and \texttt{TRACTOR}, accordingly), which is expected due to the fact that larger bars reside in larger discs~\citep{Erwin_2019}. Moreover, we find that the ray size correlates closely with all geometric properties such as various disc sizes, including the vertical scale $z_0$ (and all sizes from panel (c) in Fig.~\ref{fig:sampleprops4}) and are closely anti-correlated with all magnitudes measurements collected in Sec.~\ref{sec:samplepprops}. As an additional test, we also tried to use the maximal value of the ray sizes instead of averaging them, but this almost insignificantly shifts the distributions towards higher values and, thus, does not change our conclusions.
\par
Next, we compared the results obtained with those available in literature. X-structures measured by~\cite{Savchenko_etal2017} in 22 galaxies show a scatter of $\mathrm{ray}/h$ values in the range from about 0.25 to 2.1 with the mean value  $\approx 1.2$ (see figure 5 in this work and filled bins in Fig.~\ref{fig:ray_to_h_EGIS}). \cite{Smirnov_Sotnikova2018} calculated the sizes of X-strictures from photometric cuts without using the decomposition procedure. Thus, the comparison may not be entirely correct, but still interesting. For their $N$-body models, the sizes of X-structures are on average smaller than in~\cite{Savchenko_etal2017}, and sizes themselves are almost equal to one unit of the disc scale. The difference may be due to the fact that~\cite{Smirnov_Sotnikova2018} actually distinguished different scales in their discs (inner and outer) and scaled the sizes of the X-structures to the outer scale, which is usually larger than the inner scale.
\par
However, in both of the aforementioned works, the measured values of $\mathrm{ray}/h$ are substantially larger than we find. The reason for this discrepancy can be found using the photometric model of the X-shaped bulge introduced in \citet{Smirnov_Savchenko2020}. This model allows one to measure the ray size using a certain characteristic scale $r_\mathrm{X}$, which is analogous to the effective radius of the Sersic profile. In \citet{Savchenko_etal2017}, the rays were measured to the point where the intensity fell below a certain level, which depended on the background and roughly corresponded to the surface brightness level of 24~mag/arcsec$^2$ in the $r$ band. On the contrary, in this work we judge their lengths only by eye, which can introduce some systematics. Fortunately, we have in sample several galaxies measured in \citet{Savchenko_etal2017} and \citet{Smirnov_Savchenko2020}, and we can compare the ray differences in terms of $r_\mathrm{X}$. The results are shown in the inner subplot of Fig.~\ref{fig:ray_to_h_EGIS}. We found that, on average, the rays measured to absolute level contain \mbox{$(2-3) r_\mathrm{X}$}, while  the same objects, but measured by eye, have a shorter length, about $1.5 r_\mathrm{X}$. This is natural given how the human eye reacts to exponential contrast changes. Taking into account this result, we can conclude that for a correct comparison with previous works, it is necessary to multiply the obtained values of $\mathrm{ray}/h$ by factor close to 2. It is easy to see that, regardless of the general roughness of the method used, the resulting distribution after correction will be close to the carefully measured X-structure sizes in \citet{Savchenko_etal2017}, which once again confirms the reliability of the created sample.

\subsection{B/PS bulges frequencies and stellar masses}
\label{sec:stellarmasses}

In \citet{Erwin_Debattista2017} for 84 local galaxies and later in \citet{Kruk_etal2019} for more distant galaxies, the authors argued that an observed B/PS bulge fraction increases significantly for galaxies with stellar masses $\log M_{\star} \gtrsim 10.4$. The result was found for galaxies with moderate inclinations and is observed for both local and distant objects up $z \approx 0.5-1.0$. As far as we know, this result has never been tested for edge-on galaxies. To do this, we found the intersection for our sample and galaxies from \citet{Kauffmann2003}, where the stellar masses were estimated based on two indices of  stellar absorption lines. We found 483 (52:225:206 of each quality class accordingly) objects at the intersection and used the median dust-corrected mass \texttt{rml50} as recommended by the authors. Then, following the \cite{Kruk_etal2019} method, who used the same data source for masses, we divided the galaxies in bins by their masses and estimated the relative frequency of B/PS bulges to the total number of galaxies from \citet{Kauffmann2003} Stellar Mass Catalogue DR4 in every bin.
\par
\begin{figure}
\includegraphics[width=0.95\columnwidth]{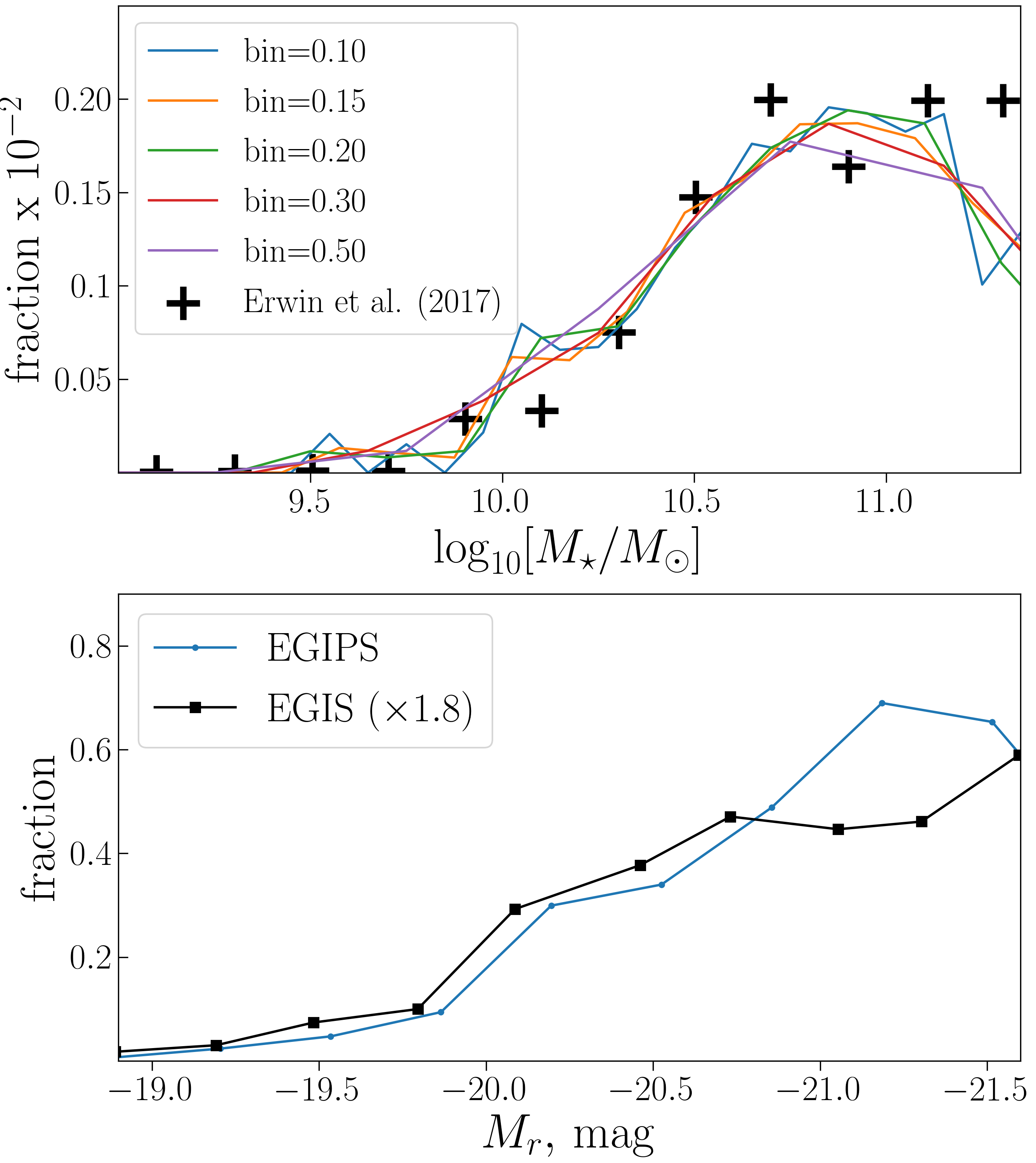}
\caption{Frequencies of stellar mass $\log M$ bins for $\approx 500$ of our galaxies. The masses were taken from \citet{Kauffmann2003}, then we divided the number of galaxies with B/PS bulges from our sample in mass bin by total number in that  bin (see also figure~10 in \citealt{Kruk_etal2019}). Lines of different colours represent different bin sizes. The crosses show the frequencies from \citet{Erwin_Debattista2017} multiplied by 0.002 for better visibility. The bottom plot shows the frequencies for our galaxies by absolute stellar magnitude $M_r$ in the $r$ band for EGIS and EGIPS distributions, calculated using the same method as in Sec.~\ref{sec:samplepprops}.}
\label{fig:Ms_freq}
\end{figure}

The results obtained for various bin sizes are presented in Fig.~\ref{fig:Ms_freq}. The figure clearly shows the previously observed increase in the fraction of B/PS bulges for massive galaxies with $\log M_{\star} \gtrsim 10.4$. This curve is very similar to that obtained in \citet{Erwin_Debattista2017} as shown by the superimposed dots in the figure. The effect is the same for different bin sizes and does not depend on the selected B/PS bulge quality class. As a sanity check, we also test what frequencies we will obtain if we plot the same graph for all EGIS edge-on galaxies (704 in the intersection with \citet{Kauffmann2003}, 45 with B/PS). As expected, we found an almost uniform distribution around 0.002, i.e. edge-on galaxies are equally likely to be found in each mass bin.  It is also interesting that in our sample there is only one outlier of the galaxy with a very low mass estimate in \citet{Kauffmann2003} equal to $\log M_{\star} \approx 8.2$, but with clearly visible B/PS bulge, i.e. EON\_179.631\_43.947 (NGC4013). Since other sources estimate the mass $\log M_{\star} > 10$ based on the data in the $3.6\mu\rm{m}$ band, and the galaxy is clearly massive, we assume that in this case the problem is related to the \citet{Kauffmann2003} data.
\par
From the confirmed mass dependence, one can expect to obtain a similar result for the luminosity fractions of galaxies. We checked such frequencies using the absolute stellar magnitudes $M_r$ in the $r$ band. Although depending on distance and extinction estimates, using $M_r$ should be more accurate as it is less model dependent than mass. The resulting frequencies, calculated for the entire EGIS sample and the subsample of galaxies with B/PS bulges, are shown in Fig.~\ref{fig:Ms_freq} in the bottom panel, where the distribution of $M_r$ for the entire EGIS sample can be found in figure~10 from \citet{Bizyaev_etal2014}. It is clear that we indeed see a certain analogue of the mass curve, and the fraction of galaxies with B/PS bulges increases significantly after $M_r < -19.5$~mag. We also tried to estimate and show on the same figure the frequencies calculated using the entire EGIPS sample, which demonstrate good agreement with the EGIS data. It is important to note here that in this case the exact values of the fraction are slightly higher than might be expected, because the SDSS data required to estimate $M_r$ is only available for half of the 13 thousands galaxies used to search for B/PS candidates (see Sec.~\ref{sec:data_description}).

\section{Conclusions}
\label{sec:conclusion}

In this work, we present the largest up-to-date sample of 1925 edge-on or near edge-on galaxies that have B/PS bulges and X-structures. The galaxies were searched among objects of the Catalog of Edge-on disc Galaxies from SDSS (EGIS)~\citep{Bizyaev_etal2014} and the Edge-on Galaxies in the Pan-STARRS survey (EGIPS)~\citet{2022MNRAS.tmp..256M} using images from the Dark Energy Spectroscopic Instrument (DESI) Legacy Imaging Survey~\citep{Dey2018}. The identification itself was based on the characteristic X-shaped features observed in the residual images (the original minus model) of the corresponding galaxies. 
\par
The sample was prepared in four main steps:
\begin{enumerate}
    \item The residuals of all EGIS galaxies were analysed manually to establish the presence of X-structures. Based on this, the entire sample of EGIS galaxies was divided into two subsamples: with and without X-structures.
    \item Both samples were then used to train a suitable Artificial Neural Network (ANN) that can automatically distinguish X-shaped features in a large sample of real galaxies. 
    \item The trained ANN was applied to the entire sample of EGIPS galaxies. As a result, a preliminary sample was built, consisting of 2123 objects.
    \item The preliminary sample was then manually examined in the framework of the Zooniverse project. Each of the authors of this paper was asked to answer several questions regarding the strength and general appearance of the X-structure and, in addition, the dust component. 198 objects were discarded after this step, while all others, where the X-shape was indicated, were divided into three quality classes, namely ``1'', ``2'', and ``3'', depending on the strength of the X-structure (class ``1'' has the strongest and largest X-structures, and class ``3'' has the weakest and shortest).
\end{enumerate}
After all these steps were completed, 
we obtained the largest up-to-date sample of 1925 galaxies with X-structures and B/PS bulges.
\par
To validate our sample, we performed various tests, including cross-validation with samples from other works, as well as analysis of the distributions over distances, stellar magnitudes in different bands, colours, and sizes. We found a good agreement between our quality classes and those adopted by~\cite{Yoshino_Yamauchi2015}. Comparing our sample and the results of~\cite{zoodeca} for B/PS bulges, which were obtained based on a large collection votes from the Galaxy Zoo project, we found that the latter tend to significantly underestimate the number of B/PS bulges observed. For a reasonable number of votes per object, we see only a few dozens of B/PS bulges from tens of thousands edge-on galaxies. This fact additionally shows how difficult the task of B/PS bulges detection is and emphasise the value of the obtained sample in this work.
\par
For distances and stellar magnitudes, we have shown that they are distributed in a typical manner, with an average distance and absolute magnitude in the $g$ band of about~200~Mpc and -20.5~mag, respectively. Studying the corresponding distributions for individual classes, we found physically consistent trends, that is, class ``1'' galaxies are brighter and closer, while galaxies of classes ``2'' and ``3'' are fainter and more distant. 
For the morphological types, we confirmed that our sample has a peak at Sab-Sb types and a sharp decline at Scd types. This is almost the same as seen in~\cite{Erwin_Debattista2017}, but not in~\cite{Yoshino_Yamauchi2015}, where Sc types were more numerous. 
\par

\par
\par
Based on the participants` entries, made during the voting in the Zoouniverse project, we have identified a number of interesting galaxies, in which the B/PS bulges and X-structures have either unusual morphology or are simply very strong, and, nevertheless, their B/PS bulges were not previously investigated (except NGC~522). Such galaxies are promising candidates for future decomposition studies.
\par
Using publicly available decomposition data from \citet{Bizyaev_etal2014}, \citet{sdss2011}, and \citet{Yoshino_Yamauchi2015}, we compared typical $B/D$ ratios of galaxies with and without X-structures. 
In each case, we found that X-shaped galaxies have, on average, higher values of $B/D$ than non X-shaped galaxies, and that in all works galaxies of ``1'' quality class demonstrate higher $B/D$ than other two. At the same time, it was found that for galaxies with B/PS bulges the typical value itself varies from study to study as follows: $\approx 0.3$ for~\cite{Bizyaev_etal2014}, $\approx 0.5$ for~\cite{sdss2011}, and $\approx 0.2$ for~\cite{Yoshino_Yamauchi2015}.
\par
Using the Zooniverse benchmark, we studied how the opening angles of the X-structure (an angle between the disc plane and the rays of the X-structures) and the length of the ray are distributed for our galaxies. These values for each galaxy were derived based on the participants' answers to a request to identify the X-structure ray in the residue image. For class ``1'', the obtained distribution of angles covers the range from 25 $\deg$ to 45 $\deg$ and has a peak of about 35 $\deg$. For classes ``2'' and ``3'', the distributions are wider with the same minimal value but with a larger maximum of 50 $\deg$. The observed range of values is in a good agreement with the predictions of $N$-body simulations by~\cite{Smirnov_Sotnikova2018}, where the minimal value of the opening angle turned out to be 25 $\deg$ in models with a significant  contribution of dark matter to the total gravitational potential.
\par
For rays sizes, we found that they are usually smaller than for galaxies in the sample by~\cite{Savchenko_etal2017}, with a typical value of about 0.6 of the exponential disc scale. Comparing the sizes measured in~\cite{Savchenko_etal2017}, ~\cite{Smirnov_Savchenko2020}, and in this work, we came to the conclusion that the sizes measured in~\cite{Savchenko_etal2017} and this work should differ by about two times due to differences in approaches to measuring the length of the ray. If we multiply the sizes measured in the present work by the mentioned factor, they become compatible with the sizes of~\cite{Savchenko_etal2017}.
\par 
In previous works~\citep{Erwin_Debattista2017, Li_etal2017,Kruk_etal2019}, a sharp increase in B/PS bulge frequencies for massive galaxies ($\log M_{\star} \gtrsim 10.4$) was found. These results were obtained for galaxies of different inclinations ($i \lesssim 70~\deg$) and are based on the theoretical assumption that the boxy shape of isophotes, as well as the presence of barlenses, are indicators of the presence of a B/PS bulge in a particular galaxy. Our sample allowed us to test whether these results are valid for edge-on galaxies, where B/PS bulges can be identified directly without the need for additional theoretical input. For our sample, we found a steep increase in B/PS bulge frequencies at~$\log M_{\star} \gtrsim 10.4$, which is very similar to what was obtained in previous works. This strengthens the previously obtained results and, at the same, additionally validates the method for identifying galaxies with B/PS bulges based by the  appearance of the X-shape in the residue image.
\par
Concluding this work, we would like to note the following. 
B/PS bulges are important components of the secular evolution of galaxies. Studying them systematically is critical to understanding various aspects of bar physics. Our sample was created for this very purpose, and it is difficult to overestimate how valuable it is for this field to have such a large and representative sample. In future, we plan to conduct a more detailed decomposition study of our galaxies in order to provide more constraints on how the bars acquire their vertical structure in real galaxies, and to test the various relationships between the B/PS bulges and the properties of their host galaxies, established in \mbox{$N$-body} simulations. In addition, it is important that our trained ANN can be applied again to new data in order to increase the sample size if, for example, a larger sample of edge-on galaxies appears.
\section*{Acknowledgements}
We acknowledge financial support from the grant of the Russian Foundation for Basic Researches number 19-02-00249.
\par
We thank the anonymous referee for his/her review and appreciate the comments, which contributed to improving the quality of the article. We thank Dmitry Makarov and all researches involved in Edge-on Galaxies in the Pan-STARRS survey (EGIPS) database creation.
\par
\par
The Legacy Surveys consist of three individual and complementary projects: the Dark Energy Camera Legacy Survey (DECaLS; Proposal ID \#2014B-0404; PIs: David Schlegel and Arjun Dey), the Beijing-Arizona Sky Survey (BASS; NOAO Prop. ID \#2015A-0801; PIs: Zhou Xu and Xiaohui Fan), and the Mayall z-band Legacy Survey (MzLS; Prop. ID \#2016A-0453; PI: Arjun Dey). DECaLS, BASS and MzLS together include data obtained, respectively, at the Blanco telescope, Cerro Tololo Inter-American Observatory, NSF’s NOIRLab; the Bok telescope, Steward Observatory, University of Arizona; and the Mayall telescope, Kitt Peak National Observatory, NOIRLab. The Legacy Surveys project is honored to be permitted to conduct astronomical research on Iolkam Du’ag (Kitt Peak), a mountain with particular significance to the Tohono O’odham Nation. NOIRLab is operated by the Association of Universities for Research in Astronomy (AURA) under a cooperative agreement with the National Science Foundation. This project used data obtained with the Dark Energy Camera (DECam), which was constructed by the Dark Energy Survey (DES) collaboration. Funding for the DES Projects has been provided by the U.S. Department of Energy, the U.S. National Science Foundation, the Ministry of Science and Education of Spain, the Science and Technology Facilities Council of the United Kingdom, the Higher Education Funding Council for England, the National Center for Supercomputing Applications at the University of Illinois at Urbana-Champaign, the Kavli Institute of Cosmological Physics at the University of Chicago, Center for Cosmology and Astro-Particle Physics at the Ohio State University, the Mitchell Institute for Fundamental Physics and Astronomy at Texas A\&M University, Financiadora de Estudos e Projetos, Fundacao Carlos Chagas Filho de Amparo, Financiadora de Estudos e Projetos, Fundacao Carlos Chagas Filho de Amparo a Pesquisa do Estado do Rio de Janeiro, Conselho Nacional de Desenvolvimento Cientifico e Tecnologico and the Ministerio da Ciencia, Tecnologia e Inovacao, the Deutsche Forschungsgemeinschaft and the Collaborating Institutions in the Dark Energy Survey. The Collaborating Institutions are Argonne National Laboratory, the University of California at Santa Cruz, the University of Cambridge, Centro de Investigaciones Energeticas, Medioambientales y Tecnologicas-Madrid, the University of Chicago, University College London, the DES-Brazil Consortium, the University of Edinburgh, the Eidgenossische Technische Hochschule (ETH) Zurich, Fermi National Accelerator Laboratory, the University of Illinois at Urbana-Champaign, the Institut de Ciencies de l'Espai (IEEC/CSIC), the Institut de Fisica d'Altes Energies, Lawrence Berkeley National Laboratory, the Ludwig Maximilians Universitat Munchen and the associated Excellence Cluster Universe, the University of Michigan, NSF's NOIRLab, the University of Nottingham, the Ohio State University, the University of Pennsylvania, the University of Portsmouth, SLAC National Accelerator Laboratory, Stanford University, the University of Sussex, and Texas A\&M University. BASS is a key project of the Telescope Access Program (TAP), which has been funded by the National Astronomical Observatories of China, the Chinese Academy of Sciences (the Strategic Priority Research Program “The Emergence of Cosmological Structures” Grant \# XDB09000000), and the Special Fund for Astronomy from the Ministry of Finance. The BASS is also supported by the External Cooperation Program of Chinese Academy of Sciences (Grant \# 114A11KYSB20160057), and Chinese National Natural Science Foundation (Grant \# 11433005). The Legacy Survey team makes use of data products from the Near-Earth Object Wide-field Infrared Survey Explorer (NEOWISE), which is a project of the Jet Propulsion Laboratory/California Institute of Technology. NEOWISE is funded by the National Aeronautics and Space Administration. The Legacy Surveys imaging of the DESI footprint is supported by the Director, Office of Science, Office of High Energy Physics of the U.S. Department of Energy under Contract No. DE-AC02-05CH1123, by the National Energy Research Scientific Computing Center, a DOE Office of Science User Facility under the same contract; and by the U.S. National Science Foundation, Division of Astronomical Sciences under Contract No. AST-0950945 to NOAO.
\par
Funding for the Sloan Digital Sky 
Survey IV has been provided by the 
Alfred P. Sloan Foundation, the U.S. 
Department of Energy Office of 
Science, and the Participating 
Institutions. SDSS-IV acknowledges support and 
resources from the Center for High 
Performance Computing  at the 
University of Utah. The SDSS 
website is www.sdss.org. 
SDSS-IV is managed by the 
Astrophysical Research Consortium 
for the Participating Institutions 
of the SDSS Collaboration including 
the Brazilian Participation Group, 
the Carnegie Institution for Science, 
Carnegie Mellon University, Center for 
Astrophysics | Harvard \& 
Smithsonian, the Chilean Participation 
Group, the French Participation Group, 
Instituto de Astrof\'isica de 
Canarias, The Johns Hopkins 
University, Kavli Institute for the 
Physics and Mathematics of the 
Universe (IPMU) / University of 
Tokyo, the Korean Participation Group, 
Lawrence Berkeley National Laboratory, 
Leibniz Institut f\"ur Astrophysik 
Potsdam (AIP),  Max-Planck-Institut 
f\"ur Astronomie (MPIA Heidelberg), 
Max-Planck-Institut f\"ur 
Astrophysik (MPA Garching), 
Max-Planck-Institut f\"ur 
Extraterrestrische Physik (MPE), 
National Astronomical Observatories of 
China, New Mexico State University, 
New York University, University of 
Notre Dame, Observat\'ario 
Nacional / MCTI, The Ohio State 
University, Pennsylvania State 
University, Shanghai 
Astronomical Observatory, United 
Kingdom Participation Group, 
Universidad Nacional Aut\'onoma 
de M\'exico, University of Arizona, 
University of Colorado Boulder, 
University of Oxford, University of 
Portsmouth, University of Utah, 
University of Virginia, University 
of Washington, University of 
Wisconsin, Vanderbilt University, 
and Yale University.
\par
This research made use of Astropy, a community developed core Python package for Astronomy
(Astropy Collaboration et al. 2013; Price-Whelan et al. 2018).

\section*{Data availability}
The data underlying this article will be shared on reasonable request to the corresponding author.

\bibliographystyle{mnras}
\bibliography{main}

\end{document}